\documentclass[twocolumn]{aastex62}

\begin{document}

\title{\bf The Distribution of Metallicities in the Local Galactic Interstellar Medium\footnote{Based in part on observations made with the NASA/ESA Hubble Space Telescope, obtained from the MAST data archive at the Space Telescope Science Institute. STScI is operated by the Association of Universities for Research in Astronomy, Inc., under NASA contract NAS5-26555. Based in part on data obtained from the ESO Science Archive Facility with DOI: \url{https://doi.org/10.18727/archive/50}.}}

\correspondingauthor{Adam M.~Ritchey}

\author{Adam M.~Ritchey}
\affiliation{Eureka Scientific, 2452 Delmer Street, Suite 100, Oakland, CA 96402, USA}
\email{ritchey.astro@gmail.com}

\author{Edward B.~Jenkins}
\affiliation{Department of Astrophysical Sciences, Princeton University, Princeton, NJ 08544, USA}

\author{J.~Michael Shull}
\affiliation{Department of Astrophysical and Planetary Sciences and CASA, University of Colorado, Boulder, CO 80309, USA}

\author{Blair D.~Savage}
\affiliation{Department of Astronomy, University of Wisconsin-Madison, Madison, WI 53706, USA}
\altaffiliation{Deceased.}

\author{S.~R.~Federman}
\affiliation{Department of Physics and Astronomy, University of Toledo, Toledo, OH 43606, USA}

\author{David~L.~Lambert}
\affiliation{W.~J.~McDonald Observatory and Department of Astronomy, University of Texas at Austin, Austin, TX 78712, USA}

\begin{abstract}
In this investigation, we present an analysis of the metallicity distribution that pertains to neutral gas in the local Galactic interstellar medium (ISM). We derive relative ISM metallicities for a sample of 84 sight lines probing diffuse atomic and molecular gas within 4 kpc of the Sun. Our analysis is based, in large part, on column density measurements reported in the literature for 22 different elements that are commonly studied in interstellar clouds. We supplement the literature data with new column density determinations for certain key elements and for several individual sight lines important to our analysis. Our methodology involves comparing the relative gas-phase abundances of many different elements for a given sight line to simultaneously determine the strength of dust depletion in that direction and the overall metallicity offset. We find that many sight lines probe multiple distinct gas regions with different depletion properties, which complicates the metallicity analysis. Nevertheless, our results provide clear evidence that the dispersion in the metallicities of neutral interstellar clouds in the solar neighborhood is small ($\sim$0.10 dex) and only slightly larger than the typical measurement uncertainties. We find no evidence for the existence of very low metallicity gas (as has recently been reported by De Cia et al.) along any of the 84 sight lines in our sample. Our results are consistent with a local Galactic ISM that is well mixed and chemically homogeneous.
\end{abstract}


\section{INTRODUCTION\label{sec:intro}}
Most models of Galactic chemical evolution assume that the gas enriched in the products of stellar nucleosynthesis is instantaneously mixed into the interstellar medium \citep[ISM; e.g.,][]{t95,c97}. The instantaneous mixing assumption, while not entirely realistic, can nevertheless yield model predictions that succeed in reproducing the evolutionary abundance trends of the elements \citep[e.g.,][]{f04}. There is strong observational evidence that supports the idea that the ISM, at a given Galactocentric distance, is well mixed and chemically homogeneous. In a comprehensive study of abundances in early B-type stars in the solar neighborhood, \citet{np12} found very narrow distributions in the abundances of He, C, N, O, Ne, Mg, Si, and Fe, with standard deviations equal to 0.05 dex or less. \citet{l18} conducted a thorough examination of abundances in hundreds of classical Cepheid stars. For Cepheids located within 1 kpc of the Sun, \citet{l18} finds that the standard deviation in the values of [Fe/H] is 0.10 dex.

\citet{ac21} presented results on radial abundance gradients in the Milky Way from a uniform analysis of 42 Galactic H~{\sc ii} regions. They find that the dispersion in the values of O/H and N/H with respect to the computed gradients is 0.07 and 0.11 dex, respectively. \citet{r18} reported on the gas-phase abundances of O and several neutron-capture elements for a sample of 128 sight lines probing the neutral ISM in the solar vicinity. The analysis of elemental abundances in the neutral ISM is complicated by the presence of significant depletion onto dust grains for most elements \citep[e.g.,][]{j09}. Adopting the methodology of \citet{j09}, \citet{r18} identified and modelled the depletion behaviors for the elements they considered. For O, Ge, and Kr, which have the largest number of abundance measurements in the sample, \citet{r18} find that the dispersions in the abundances with respect to the depletion trends are 0.07, 0.10, and 0.07 dex, respectively. Numerous other studies have examined the homogeneity of gas-phase interstellar abundances in the solar neighborhood \citep[e.g.,][]{c96,m97,m98,c03,c04,c06}. These studies have generally found only small variations in the gas-phase abundances once the effects of dust-grain depletion are taken into account.

From a theoretical perspective, the Galactic ISM is expected to be well mixed in the azimuthal direction \citep[e.g.,][]{e75,rk95}. Turbulent transport in the shear flow of differential rotation should reduce azimuthal inhomogeneities in large disk galaxies like the Milky Way within a characteristic timescale of $10^9$ yr \citep{rk95}. On smaller scales, turbulent mixing due to supernova explosions \citep{dm02} and thermal instabilities \citep{yk12} leads to even shorter mixing timescales. \citet{p15} studied turbulent mixing driven by gravitational instabilities and found that metallicity inhomogeneities are destroyed in the azimuthal direction in less than a galactic orbital period.

Recently, however, \citet{dc21} reported finding large metallicity variations in the neutral ISM toward stars located within 3 kpc of the Sun. After applying two separate techniques to account for dust-grain depletion, \citet{dc21} report depletion-corrected metallicities in the range $-0.76\le[{\rm M}/{\rm H}]\le+0.26$ dex for their sample of 25 sight lines. They report a mean value of $[{\rm M}/{\rm H}]$ of $-0.26$ dex and a standard deviation of 0.28 dex. \citet{dc21} attribute the low metallicity material to the effects of metal-poor infall from high-velocity clouds combined with inefficient mixing in the Galactic disk. However, these conclusions run counter to the growing body of evidence from observational and theoretical work that the ISM, at a given Galactocentric distance, is (and is expected to be) well mixed and chemically homogeneous.

Upon closer examination, we find that there are serious flaws in the analysis techniques adopted in \citet{dc21} that undermine their conclusions. For example, they derive column densities from relatively strong interstellar absorption lines (e.g., Zn~{\sc ii} $\lambda\lambda2026,2062$) using spectra obtained at moderate resolution ($\Delta v\sim10$ km~s$^{-1}$). While attempts are made to account for unresolved saturation in the absorption profiles, it appears that these attempts are inadequate in some cases and that many of the Zn~{\sc ii} column densities in particular may be underestimated. A more serious issue is that the elements used by \citet{dc21} for their metallicity estimates (Si, Ti, Cr, Fe, Ni, and Zn) are the more refractory elements, which are significantly impacted by dust depletion. Volatile elements (such as C, N, O, and Kr), which are more reliable metallicity indicators, are not included in their analysis, even though they yield conflicting results.

In this investigation, we seek to derive, in a much more definitive way, the metallicity distribution for the local Galactic ISM. Our analysis relies mainly on high-quality gas-phase abundance measurements reported in the literature. However, we also obtain new column density determinations for certain key elements and for several important and/or interesting sight lines  in order to fill in some of the gaps in the literature data. The methodology used in our metallicity analysis, which is based on the unified representation of interstellar depletions devised by \citet{j09}, is described in Section~\ref{sec:method}. The sources and methods used to obtain column density measurements for our sample are described in Section~\ref{sec:sample}. The metallicity estimates themselves are provided in Section~\ref{sec:metallicities}. In Section~\ref{subsec:comparison}, we present a detailed comparison between our metallicity determinations and those of \citet{dc21}. In Section~\ref{subsec:variations}, we discuss the implications of our results in the context of other measurements of metallicity variations in the Galactic disk. We summarize our conclusions in Section~\ref{sec:conclusions}. An appendix provides a compilation of all of the column density measurements used in the metallicity analysis.

\section{METHODOLOGY\label{sec:method}}
In a comprehensive analysis of interstellar depletions, \citet{j09} utilized measurements of elemental abundances reported in the literature for a sample of 243 sight lines to examine the depletion characteristics of 17 different elements. \citet{j09} developed a unified representation of gas-phase element depletions predicated on the empirical observation that, while different elements exhibit different degrees of depletion, the depletions of most elements tend to increase in a systematic way as the overall strength of depletions increases from one line of sight to the next. Sight lines with stronger depletions are thought to contain higher proportions of denser and/or colder gas \citep{s85,ss96b}, while sight lines with very low depletions may contain grains that have been partially destroyed through sputtering in interstellar shocks \citep[e.g.,][]{s77}. The increase in the overall strength of depletions, from low-density diffuse atomic gas to higher-density diffuse molecular clouds, may then be regarded as evidence of grain growth in the ISM \citep{ds80,j86,ss96b,j09,j13}. Many grain history calculations have suggested that the survival of dust grains against destruction via shock sputtering requires continued growth in diffuse clouds \citep{ss83,j94,s15}. With these considerations, \citet{j09} defined a sight-line depletion strength factor, denoted $F_*$, which indicates the extent to which grain growth has progressed in the gas sampled by a particular line of sight. The value of $F_*$ for any given sight line is based on a weighted average of the available observed depletions for that direction \citep[see Equation (4) in][]{j09}. Sight lines showing strong depletions, such as those seen in the low velocity ($v_{\rm LSR}=-1$~km~s$^{-1}$) component toward $\zeta$~Oph \citep[e.g.,][]{s92}, have depletion strengths near $F_*=1$, while those with the weakest depletions have $F_*$ values close to zero.

The depletion of an element $X$ from the gas phase is defined with respect to an adopted cosmic reference abundance, typically the solar system abundance of the element. In logarithmic terms, we have:

\begin{equation}
[X/{\rm H}]=\log N(X)-\log N({\rm H}_{\rm tot})-\log(X/{\rm H})_{\sun},
\end{equation}

\noindent
where $N(X)$ is the total column density of element $X$ in its dominant ionization stage, $N({\rm H}_{\rm tot})=N({\rm H~\textsc{i}})+2N({\rm H}_2)$ is the total hydrogen column density along the line of sight, and $(X/{\rm H})_{\sun}$ is the solar reference abundance. In the framework developed by \citet{j09}, the logarithmic depletion of an element depends on the sight-line depletion strength factor according to:

\begin{equation}
[X/{\rm H}]=B_X+A_X(F_*-z_X),
\end{equation}

\noindent
where the depletion parameters $A_X$, $B_X$, and $z_X$ are unique to each specific element. As described in more detail in \citet{j09}, the slope parameter $A_X$ indicates how quickly the depletion of a particular element strengthens as the growth of dust grains progresses within interstellar clouds \citep[see also][]{j13}. The intercept parameter $B_X$ indicates the expected depletion of element $X$ at $F_*=z_X$, where $z_X$ represents a weighted mean value of $F_*$ for the particular set of sight lines with depletion measurements available for that element. Values of the coefficients $A_X$ and $B_X$ are obtained for a specific element through the evaluation of a least-squares linear fit, with $[X/{\rm H}]$ as the dependent variable and $F_*$ the independent variable. (The reason for the additional term involving $z_X$ in Equation (2) is that, for a particular choice of $z_X$, there is a near zero covariance between the formal fitting errors for the solutions of $A_X$ and $B_X$ \citep[see][]{j09}.)

\citet{j09} obtained values of the element-specific depletion parameters for 17 different elements: C, N, O, Mg, Si, P, S, Cl, Ti, Cr, Mn, Fe, Ni, Cu, Zn, Ge, and Kr. Later, \citet{r18}, adopting the \citet{j09} methodology, derived depletion parameters for B, Ga, As, Cd, Sn, and Pb and re-evaluated the depletion parameters for O, Ge, and Kr using larger samples of elemental abundance measurements. Most recently, \citet{r23} re-determined the values of the depletion parameters for the elements P and Cl, both to account for recent updates in the oscillator strengths of the relevant P~{\sc ii} and Cl~{\sc i} transitions and to correct for a deficiency in the Cl analysis presented in \citet{j09}. In the original analysis, \citet{j09} considered only Cl~{\sc ii} column densities in deriving total Cl abundances. However, Cl is unique among most other elements commonly studied in the ISM in that either its neutral or its singly-ionized form can dominate the total abundance depending on the amount of molecular hydrogen present along the line of sight \citep[e.g.,][]{jy78}. \citet{r23} included column densities of both Cl~{\sc i} and Cl~{\sc ii} in their derivations of total Cl abundances and Cl depletion factors.

Based on the formalism developed to study gas-phase element depletions, \citet{j09} proposed a method that could be used to examine the metallicities of distant (extragalactic) absorption systems, such as those seen in the optical spectra of high-redshift quasars. As we show below, the metallicity determined according to this method is the metallicity of the absorption system relative to the average metallicity that characterizes the local Galactic ISM. (Hereafter, we refer to this quantity as the ``relative ISM metallicity''.) In the present work, we use this method to examine the spread in the metallicities determined for sight lines probing the solar neighborhood (out to $\sim$4~kpc).

To determine the relative ISM metallicity for a given sight line, we first equate the two expressions for $[X/{\rm H}]$ presented in Equations (1) and (2):

\begin{equation}
\log N(X)-\log N({\rm H}_{\rm tot})-\log(X/{\rm H})_{\sun}=B_X+A_X(F_*-z_X).
\end{equation}

\noindent
Rearranging terms, we have:

\begin{equation}
\log N(X)-\log(X/{\rm H})_{\sun}-B_X+A_Xz_X=\log N({\rm H}_{\rm tot})+F_*A_X.
\end{equation}

\noindent
The quantity $N({\rm H}_{\rm tot})$ appearing on the righthand side of Equation (4) can be interpreted as the total hydrogen column density that would be expected based on the column density of element $X$ and the overall strength of depletions along the line of sight ($F_*$). However, this expectation holds only if the metallicity of the gas is equal to the average ISM metallicity in the solar neighborhood. If the metallicity is somewhat different, then the true hydrogen column density will be higher or lower than the predicted value. Thus, if the total hydrogen column density is known independently from observations, we can rewrite Equation (4) in terms of the relative ISM metallicity:

\begin{equation}
[{\rm M}/{\rm H}]_{\rm ISM}=\log N({\rm H}_{\rm tot})_{\rm pred}-\log N({\rm H}_{\rm tot})_{\rm obs}.
\end{equation}

\noindent
Subtracting $\log N({\rm H}_{\rm tot})_{\rm obs}$ from both sides of Equation (4) yields:

\begin{equation}
[X/{\rm H}]_{\rm obs}-B_X+A_Xz_X=[{\rm M}/{\rm H}]_{\rm ISM}+F_*A_X,
\end{equation}

\noindent
where

\begin{equation}
[X/{\rm H}]_{\rm obs}=\log N(X)-\log N({\rm H}_{\rm tot})_{\rm obs}-\log(X/{\rm H})_{\sun}
\end{equation}

\noindent
is the observed depletion of element $X$. Since the quantities $A_X$, $B_X$, and $z_X$ have been tabulated for many different elements, a least-squares linear fit to the equation:

\begin{equation}
y=a+bx,
\end{equation}

\noindent
where

\begin{equation}
y=[X/{\rm H}]_{\rm obs}-B_X+A_Xz_X
\end{equation}

\noindent
and

\begin{equation}
x=A_X,
\end{equation}

\noindent
will yield values for the coefficients

\begin{equation}
a=[{\rm M}/{\rm H}]_{\rm ISM}
\end{equation}

\noindent
and

\begin{equation}
b=F_*.
\end{equation}

\noindent
As stated above, the metallicity derived in this way is the metallicity of the absorption system relative to the average ISM metallicity in the solar neighborhood. It is \emph{not} the metallicity relative to an adopted solar (or cosmic) abundance standard. To see why this is the case, let us consider the definition of the intercept parameter $B_X$. Rearranging Equation (2) and setting $F_*=z_X$ gives:

\begin{equation}
B_X=[X/{\rm H}]_{F_*=z_X}=\log(X/{\rm H})_{F_*=z_X}-\log(X/{\rm H})_{\sun},
\end{equation}

\noindent
where $\log(X/{\rm H})_{F_*=z_X}$ is the (average) logarithmic abundance of element $X$ evaluated at $F_*=z_X$ (according to the linear fit used to determine $A_X$ and $B_X$). Substituting the above expression for $B_X$ into Equation (6) and rearranging terms yields:

\begin{equation}
[{\rm M}/{\rm H}]_{\rm ISM}=\log(X/{\rm H})_{\rm obs}-\log(X/{\rm H})_{F_*=z_X}-A_X(F_*-z_X).
\end{equation}

\noindent
Note that Equation (14) makes no reference to any solar (or cosmic) abundance standard. The terms involving the solar abundance of element $X$ have cancelled out of the equation. Nevertheless, despite this apparent shortcoming, the method described here can still provide us with useful information on the variations in metallicity that characterize the interstellar gas in the solar vicinity. In Section~\ref{sec:metallicities}, we use this method to derive relative ISM metallicities for a sample of 84 sight lines probing the local Galactic ISM.

\section{CONSTRUCTING THE SAMPLE\label{sec:sample}}
In order to obtain a statistically significant result for the metallicity distribution in the solar neighborhood, we need high-quality column density measurements for a variety of elements in their dominant ionization stage for a relatively large sample of interstellar sight lines. Fortunately, many such measurements have been published over the past several decades. Thus, in constructing our sample, we rely primarily on column density measurements reported in the literature (Section~\ref{subsec:lit_data}). However, we supplement these data with new column density determinations for certain key elements and for several important and/or interesting sight lines (as described in Section~\ref{subsec:new_data}).

\subsection{Accumulation of Data from the Literature\label{subsec:lit_data}}
In compiling our database of high-quality column density measurements from the literature, we focus on the following elements: B, C, N, O, Mg, Si, P, Cl, Ti, Cr, Mn, Fe, Ni, Cu, Zn, Ga, Ge, As, Kr, Cd, Sn, and Pb. All 22 of these elements have been analyzed in a way consistent with the methodology devised by \citet{j09}, and all have values of the depletion parameters $A_X$, $B_X$, and $z_X$ tabulated in the literature \citep{j09,r18,r23}.\footnote{While \citet{j09} determined depletion parameters for S, he pointed out difficulties in working with that element. The only S~{\sc ii} transitions that are available are the ones in the strong triplet that spans 1250 to 1260~\AA{}. For most sight lines, these absorption features are strongly saturated. When the total hydrogen column density is low enough, the lines are not saturated, but contributions of S~{\sc ii} from H~{\sc ii} regions can lead to misleading conclusions on the abundance of S in the neutral gas.} The inclusion of as many different elements as possible in our least-squares linear fits helps to ensure that the relative ISM metallicities we derive are robust.

Most of the dominant ions of the elements listed above have electronic transitions out of the ground state with wavelengths in the UV portion of the spectrum accessible to the Hubble Space Telescope (HST). For these ions, we restricted our search to determinations of column densities based on observations obtained with the Space Telescope Imaging Spectrograph (STIS) or the Goddard High Resolution Spectrograph (GHRS). We do not include in our sample measurements made using earlier spaceborne instruments such as Copernicus or the International Ultraviolet Explorer. Abundance determinations made using these earlier UV instruments tend to be less precise than those based on HST observations \citep[e.g.,][]{m98}. For a few ions (e.g., N~{\sc i}, Cl~{\sc ii}, and Fe~{\sc ii}), many of the observed transitions have wavelengths below 1200~\AA{}, a regime covered by the Far Ultraviolet Spectroscopic Explorer (FUSE). Finally, all of the observed transitions of Ti~{\sc ii} have wavelengths above 3000~\AA{}. A variety of ground-based optical telescopes have been used to study these transitions \citep[see, e.g.,][]{wc10}.

Many investigations of elemental abundances based on observations of interstellar absorption lines fall into one of two categories. Some studies focus on one or several elements and obtain abundances for a large, diverse set of interstellar sight lines \citep[e.g.,][]{c04,c06,r11,r18,r23,j19}. Others focus on a single line of sight and derive abundances for as many atomic (and molecular) species as possible \citep[e.g.,][]{s02,s03,w20}. Both types of investigations are utilized in the present study, where our aim is to compile as many reliable elemental abundance measurements as possible for as large a sample as is feasible.

\startlongtable
\begin{deluxetable*}{lcccccccccc}
\tablecolumns{11}
\tabletypesize{\small}
\tablecaption{Stellar and Sight Line Properties for the Final Sample\label{tab:sample}}
\tablehead{ \colhead{Star} & \colhead{Name} & \colhead{Sp.~Type} & \colhead{$V$} & \colhead{$E(\bv)$} & \colhead{$l$} & \colhead{$b$} & \colhead{$d$\tablenotemark{a}} & \colhead{$z$} & \colhead{$\log N({\rm H}_{\rm tot})$} & \colhead{Ref.\tablenotemark{b}} \\
\colhead{} & \colhead{} & \colhead{} & \colhead{(mag)} & \colhead{(mag)} & \colhead{(deg)} & \colhead{(deg)} & \colhead{(kpc)} & \colhead{(kpc)} & \colhead{} & \colhead{}}
\startdata
HD~1383         & \ldots & B1II         &  7.63  &  0.47  &  119.02  &   $-0.89$  &  $2.50^{+0.18}_{-0.13}$ &  $-0.039$  &  $21.54^{+0.04}_{-0.05}$ & 1 \\
HD~12323        & \ldots & ON9.2V          &  8.92  &  0.23  &  132.91  &   $-5.87$  &  $2.44^{+0.23}_{-0.22}$ &  $-0.250$  &  $21.28^{+0.04}_{-0.04}$ & 1 \\
HD~13268        & \ldots & ON8.5IIIn        &  8.18  &  0.36  &  133.96  &   $-4.99$  &  $1.77^{+0.09}_{-0.07}$ &  $-0.154$  &  $21.44^{+0.06}_{-0.07}$ & 1 \\
HD~13745        & V354~Per & O9.7IIn      &  7.90  &  0.46  &  134.58  &   $-4.96$  &  $2.28^{+0.15}_{-0.09}$ &  $-0.197$  &  $21.46^{+0.04}_{-0.05}$ & 1 \\
HD~14434        & \ldots & O5.5Vnfp     &  8.49  &  0.48  &  135.08  &   $-3.82$  &  $2.24^{+0.10}_{-0.09}$ &  $-0.150$  &  $21.47^{+0.07}_{-0.09}$ & 2 \\
HD~15137        & \ldots & O9.5II-IIIn  &  7.86  &  0.35  &  137.46  &   $-7.58$  &  $2.05^{+0.17}_{-0.13}$ &  $-0.271$  &  $21.32^{+0.07}_{-0.08}$ & 1 \\
HD~23180        & $o$~Per & B1III        &  3.83  &  0.30  &  160.36  &  $-17.74$  &  $0.341^{+0.052}_{-0.040}$ &  $-0.104$  &  $21.19^{+0.10}_{-0.14}$ & 2 \\
HD~24190        & \ldots & B2Vn         &  7.45  &  0.30  &  160.39  &  $-15.18$  &  $0.375^{+0.005}_{-0.005}$ &  $-0.098$  &  $21.30^{+0.05}_{-0.06}$ & 2 \\
HD~24398        & $\zeta$~Per & B1Ib         &  2.85  &  0.34  &  162.29  &  $-16.69$  &  $0.264^{+0.022}_{-0.022}$ &  $-0.076$  &  $21.19^{+0.06}_{-0.08}$ & 2 \\
HD~24534        & X~Per & O9.5III      &  6.72  &  0.59  &  163.08  &  $-17.14$  &  $0.596^{+0.017}_{-0.014}$ &  $-0.176$  &  $21.34^{+0.03}_{-0.04}$ & 2 \\
HD~24912        & $\xi$~Per & O7.5IIInf    &  4.06  &  0.35  &  160.37  &  $-13.11$  &  $0.418^{+0.058}_{-0.039}$ &  $-0.095$  &  $21.29^{+0.08}_{-0.09}$ & 2 \\
HD~35149        & 23~Ori & B1Vn         &  5.00  &  0.11  &  199.16  &  $-17.86$  &  $0.575^{+0.119}_{-0.078}$ &  $-0.176$  &  $20.74^{+0.08}_{-0.10}$ & 3 \\
HD~37021        & $\theta^1$~Ori~B & B3V          &  7.96  &  0.48  &  209.01  &  $-19.38$  &  $0.375^{+0.005}_{-0.006}$ &  $-0.124$  &  $21.65^{+0.13}_{-0.19}$ & 2 \\
HD~37061        & NU~Ori & B0.5V        &  6.83  &  0.56  &  208.92  &  $-19.27$  &  $0.409^{+0.011}_{-0.010}$ &  $-0.135$  &  $21.73^{+0.09}_{-0.11}$ & 2 \\
HD~37903        & \ldots & B1.5V        &  7.83  &  0.35  &  206.85  &  $-16.54$  &  $0.394^{+0.003}_{-0.004}$ &  $-0.112$  &  $21.44^{+0.06}_{-0.07}$ & 2 \\
HD~52266        & \ldots & O9.5IIIn        &  7.23  &  0.29  &  219.13  &   $-0.68$  &  $1.35^{+0.06}_{-0.07}$ &  $-0.016$  &  $21.27^{+0.04}_{-0.04}$ & 1 \\
HD~53975        & \ldots & O7.5Vz        &  6.50  &  0.21  &  225.68  &   $-2.32$  &  $1.12^{+0.07}_{-0.06}$ &  $-0.045$  &  $21.09^{+0.04}_{-0.04}$ & 1 \\
HD~57061        & $\tau$~CMa & O9II         &  4.40  &  0.16  &  238.18  &   $-5.54$  &  $3.22^{+1.79}_{-1.41}$ &  $-0.311$  &  $20.70^{+0.04}_{-0.04}$ & 2 \\
HD~62542        & \ldots & B5V          &  8.03  &  0.35  &  255.92  &   $-9.24$  &  $0.366^{+0.002}_{-0.002}$ &  $-0.059$  &  $21.25^{+0.17}_{-0.28}$ & 4 \\
HD~63005        & \ldots & O7Vf         &  9.13  &  0.27  &  242.47  &   $-0.93$  &  $3.72^{+0.60}_{-0.50}$ &  $-0.060$  &  $21.31^{+0.03}_{-0.03}$ & 1 \\
HD~69106        & \ldots & B0.2V     &  7.13  &  0.20  &  254.52  &   $-1.33$  &  $1.42^{+0.10}_{-0.06}$ &  $-0.033$  &  $21.11^{+0.04}_{-0.04}$ & 1 \\
HD~73882        & NX~Vel & O8.5IV       &  7.19  &  0.70  &  260.18  &   $+0.64$  &  $0.737^{+0.031}_{-0.030}$ &  $+0.008$  &  $21.57^{+0.08}_{-0.09}$ & 2 \\
HD~75309        & \ldots & B1IIp        &  7.84  &  0.29  &  265.86  &   $-1.90$  &  $1.82^{+0.12}_{-0.13}$ &  $-0.060$  &  $21.19^{+0.03}_{-0.03}$ & 1 \\
HD~79186        & GX~Vel & B5Ia         &  5.00  &  0.40  &  267.36  &   $+2.25$  &  $1.81^{+0.34}_{-0.23}$ &  $+0.071$  &  $21.41^{+0.07}_{-0.08}$ & 2 \\
HD~88115        & \ldots & B1.5Iin      &  9.36  &  0.16  &  285.32  &   $-5.53$  &  $2.53^{+0.16}_{-0.17}$ &  $-0.243$  &  $21.04^{+0.06}_{-0.07}$ & 1 \\
HD~90087        & \ldots & O9.2III       &  8.92  &  0.28  &  285.16  &   $-2.13$  &  $2.19^{+0.11}_{-0.12}$ &  $-0.082$  &  $21.23^{+0.05}_{-0.05}$ & 1 \\
HD~91824        & \ldots & O7Vfz          &  8.14  &  0.24  &  285.70  &   $+0.07$  &  $1.83^{+0.08}_{-0.08}$ &  $+0.002$  &  $21.16^{+0.04}_{-0.04}$ & 1 \\
HD~91983        & \ldots & B1III        &  8.55  &  0.29  &  285.88  &   $+0.05$  &  $2.40^{+0.15}_{-0.14}$ &  $+0.002$  &  $21.22^{+0.05}_{-0.06}$ & 1 \\
HD~92554        & \ldots & O9.5IIn      & 10.15  &  0.39  &  287.60  &   $-2.02$  &  $4.04^{+0.31}_{-0.28}$ &  $-0.142$  &  $21.35^{+0.09}_{-0.11}$ & 1 \\
HD~93205        & V560~Car & O3.5V         &  7.75  &  0.38  &  287.57  &   $-0.71$  &  $2.25^{+0.12}_{-0.11}$ &  $-0.028$  &  $21.38^{+0.05}_{-0.05}$ & 1 \\
HD~93222        & \ldots & O7IIIf       &  8.10  &  0.36  &  287.74  &   $-1.02$  &  $2.41^{+0.14}_{-0.15}$ &  $-0.043$  &  $21.49^{+0.03}_{-0.03}$ & 1 \\
HD~94493        & \ldots & B1Ib         &  7.59  &  0.23  &  289.01  &   $-1.18$  &  $2.15^{+0.14}_{-0.12}$ &  $-0.044$  &  $21.18^{+0.05}_{-0.06}$ & 1 \\
HD~99857        & \ldots & B0.5Ib       &  7.49  &  0.35  &  294.78  &   $-4.94$  &  $1.80^{+0.08}_{-0.07}$ &  $-0.155$  &  $21.35^{+0.06}_{-0.07}$ & 1 \\
HD~99890        & \ldots & B0IIIn       &  9.26  &  0.24  &  291.75  &   $+4.43$  &  $2.53^{+0.18}_{-0.17}$ &  $+0.195$  &  $21.14^{+0.05}_{-0.05}$ & 1 \\
HD~104705       & DF~Cru & B0Ib         &  9.11  &  0.23  &  297.45  &   $-0.34$  &  $1.94^{+0.16}_{-0.15}$ &  $-0.011$  &  $21.21^{+0.05}_{-0.06}$ & 1 \\
HD~108639       & \ldots & B0.2III      &  8.57  &  0.37  &  300.22  &   $+1.95$  &  $1.98^{+0.10}_{-0.11}$ &  $+0.067$  &  $21.40^{+0.04}_{-0.04}$ & 1 \\
HD~114886       & \ldots & O9III       &  6.89  &  0.40  &  305.52  &   $-0.83$  &  $1.83^{+0.95}_{-0.78}$ &  $-0.026$  &  $21.41^{+0.05}_{-0.06}$ & 1 \\
HD~116781       & V967~Cen & B0IIIne      &  7.62  &  0.43  &  307.05  &   $-0.07$  &  $2.11^{+0.15}_{-0.13}$ &  $-0.002$  &  $21.27^{+0.05}_{-0.05}$ & 1 \\
HD~116852       & \ldots & O8.5II-IIIf        &  8.47  &  0.21  &  304.88  &  $-16.13$  &  $3.42^{+0.39}_{-0.32}$ &  $-0.949$  &  $21.01^{+0.04}_{-0.04}$ & 1 \\
HD~121968       & \ldots & B1V          & 10.26  &  0.07  &  333.97  &  $+55.84$  &  $3.94^{+0.91}_{-0.63}$ &  $+3.263$  &  $20.59^{+0.12}_{-0.16}$ & 2 \\
HD~122879       & \ldots & B0Ia         &  6.50  &  0.36  &  312.26  &   $+1.79$  &  $2.22^{+0.16}_{-0.13}$ &  $+0.069$  &  $21.39^{+0.05}_{-0.06}$ & 1 \\
HD~124314       & \ldots & O6IVnf        &  6.64  &  0.53  &  312.67  &   $-0.42$  &  $1.61^{+0.11}_{-0.10}$ &  $-0.012$  &  $21.49^{+0.05}_{-0.06}$ & 1 \\
HD~137595       & \ldots & B3Vn         &  7.49  &  0.25  &  336.72  &  $+18.86$  &  $0.751^{+0.021}_{-0.022}$ &  $+0.243$  &  $21.23^{+0.04}_{-0.05}$ & 1 \\
HD~141637       & 1~Sco & B2.5Vn       &  4.64  &  0.15  &  346.10  &  $+21.71$  &  $0.147^{+0.003}_{-0.003}$ &  $+0.054$  &  $21.13^{+0.10}_{-0.13}$ & 2 \\
HD~147683       & V760~Sco & B4V+B4V      &  7.05  &  0.39  &  344.86  &  $+10.09$  &  $0.290^{+0.002}_{-0.002}$ &  $+0.051$  &  $21.41^{+0.11}_{-0.14}$ & 2 \\
HD~147888       & $\rho$~Oph~D & B3V          &  6.74  &  0.47  &  353.65  &  $+17.71$  &  $0.124^{+0.006}_{-0.006}$ &  $+0.038$  &  $21.73^{+0.07}_{-0.09}$ & 1 \\
HD~147933       & $\rho$~Oph~A & B2V          &  5.05  &  0.45  &  353.69  &  $+17.69$  &  $0.137^{+0.003}_{-0.003}$ &  $+0.042$  &  $21.70^{+0.08}_{-0.10}$ & 2 \\
HD~148937       & \ldots & O6f?p         &  6.71  &  0.65  &  336.37  &   $-0.22$  &  $1.15^{+0.03}_{-0.03}$ &  $-0.004$  &  $21.60^{+0.05}_{-0.05}$ & 1 \\
HD~149404       & V918~Sco & O8.5Iabfp        &  5.52  &  0.68  &  340.54  &   $+3.01$  &  $1.30^{+0.13}_{-0.11}$ &  $+0.068$  &  $21.57^{+0.10}_{-0.13}$ & 5, 6 \\
HD~149757       & $\zeta$~Oph & O9.5IVnn        &  2.56  &  0.32  &    6.28  &  $+23.59$  &  $0.139^{+0.017}_{-0.015}$ &  $+0.055$  &  $21.15^{+0.03}_{-0.03}$ & 2 \\
HD~152590       & V1297~Sco & O7.5Vz          &  9.29  &  0.48  &  344.84  &   $+1.83$  &  $1.68^{+0.08}_{-0.06}$ &  $+0.053$  &  $21.47^{+0.06}_{-0.07}$ & 1 \\
HD~157857       & \ldots & O6.5IIf     &  7.78  &  0.43  &   12.97  &  $+13.31$  &  $2.22^{+0.20}_{-0.13}$ &  $+0.511$  &  $21.44^{+0.07}_{-0.08}$ & 2 \\
HD~165246       & \ldots & O8Vn     &  7.60  &  0.38  &   6.40  &  $-1.56$  &  $1.19^{+0.04}_{-0.05}$ &  $-0.032$  &  $21.45^{+0.03}_{-0.03}$ & 1 \\
HD~165955       & \ldots & B3Vn         &  9.59  &  0.15  &  357.41  &   $-7.43$  &  $1.47^{+0.12}_{-0.09}$ &  $-0.190$  &  $21.11^{+0.06}_{-0.07}$ & 2, 7 \\
HD~170740       & \ldots & B2IV-V       &  5.72  &  0.48  &   21.06  &   $-0.53$  &  $0.225^{+0.005}_{-0.005}$ &  $-0.002$  &  $21.43^{+0.05}_{-0.06}$ & 1 \\
HD~177989       & \ldots & B0III        &  9.34  &  0.23  &   17.81  &  $-11.88$  &  $2.41^{+0.20}_{-0.19}$ &  $-0.496$  &  $21.10^{+0.05}_{-0.05}$ & 1 \\
HD~185418       & \ldots & B0.5V        &  7.49  &  0.50  &   53.60  &   $-2.17$  &  $0.692^{+0.010}_{-0.009}$ &  $-0.026$  &  $21.41^{+0.04}_{-0.05}$ & 1 \\
HD~191877       & \ldots & B1Ib         &  6.27  &  0.21  &   61.57  &   $-6.45$  &  $1.73^{+0.11}_{-0.13}$ &  $-0.194$  &  $21.10^{+0.05}_{-0.06}$ & 1 \\
HD~192035       & RX~Cyg & B0III-IVn    &  8.22  &  0.34  &   83.33  &   $+7.76$  &  $1.65^{+0.06}_{-0.06}$ &  $+0.223$  &  $21.39^{+0.04}_{-0.05}$ & 1 \\
HD~192639       & \ldots & O7.5Iab        &  7.11  &  0.66  &   74.90  &   $+1.48$  &  $1.81^{+0.07}_{-0.06}$ &  $+0.047$  &  $21.48^{+0.07}_{-0.08}$ & 2 \\
HD~195455       & \ldots & B0.5III      &  9.20  &  0.10  &   20.27  &  $-32.14$  &  $2.35^{+0.35}_{-0.24}$ &  $-1.251$  &  $20.62^{+0.04}_{-0.04}$ & 1 \\
HD~195965       & \ldots & B0V          &  6.97  &  0.25  &   85.71  &   $+5.00$  &  $0.790^{+0.023}_{-0.025}$ &  $+0.069$  &  $21.08^{+0.04}_{-0.05}$ & 1 \\
HD~198478       & 55~Cyg & B3Ia         &  4.86  &  0.57  &   85.75  &   $+1.49$  &  $1.84^{+0.35}_{-0.22}$ &  $+0.048$  &  $21.51^{+0.16}_{-0.26}$ & 1 \\
HD~201345       & \ldots & ON9.2IV          &  7.76  &  0.15  &   78.44  &   $-9.54$  &  $1.83^{+0.15}_{-0.11}$ &  $-0.303$  &  $21.02^{+0.05}_{-0.05}$ & 1 \\
HD~202347       & \ldots & B1.5V        &  7.50  &  0.17  &   88.22  &   $-2.08$  &  $0.764^{+0.023}_{-0.019}$ &  $-0.028$  &  $20.94^{+0.07}_{-0.08}$ & 1 \\
HD~203374       & \ldots & B0IVpe        &  6.67  &  0.53  &  100.51  &   $+8.62$  &  $2.04^{+1.42}_{-0.76}$ &  $+0.306$  &  $21.40^{+0.04}_{-0.05}$ & 1 \\
HD~206267       & \ldots & O6Vf        &  5.62  &  0.53  &   99.29  &   $+3.74$  &  $0.790^{+0.172}_{-0.112}$ &  $+0.052$  &  $21.49^{+0.05}_{-0.06}$ & 1 \\
HD~206773       & \ldots & B0Vpe     &  6.87  &  0.45  &   99.80  &   $+3.62$  &  $0.888^{+0.016}_{-0.014}$ &  $+0.056$  &  $21.24^{+0.05}_{-0.06}$ & 1 \\
HD~207198       & \ldots & O8.5II    &  5.94  &  0.62  &  103.14  &   $+6.99$  &  $0.978^{+0.034}_{-0.027}$ &  $+0.119$  &  $21.50^{+0.05}_{-0.05}$ & 1 \\
HD~207308       & \ldots & B0.5V  &  7.49  &  0.53  &  103.11  &   $+6.82$  &  $0.906^{+0.017}_{-0.013}$ &  $+0.108$  &  $21.45^{+0.04}_{-0.05}$ & 1 \\
HD~207538       & \ldots & O9.7IV        &  7.30  &  0.64  &  101.60  &   $+4.67$  &  $0.830^{+0.013}_{-0.013}$ &  $+0.068$  &  $21.52^{+0.04}_{-0.05}$ & 1 \\
HD~208440       & \ldots & B1V          &  7.91  &  0.28  &  104.03  &   $+6.44$  &  $0.877^{+0.019}_{-0.018}$ &  $+0.098$  &  $21.33^{+0.05}_{-0.06}$ & 1 \\
HD~209339       & \ldots & O9.7IV         &  8.51  &  0.36  &  104.58  &   $+5.87$  &  $0.936^{+0.028}_{-0.024}$ &  $+0.096$  &  $21.27^{+0.04}_{-0.04}$ & 1 \\
HD~210809       & \ldots & O9Iab        &  7.56  &  0.31  &   99.85  &   $-3.13$  &  $3.66^{+0.52}_{-0.34}$ &  $-0.200$  &  $21.35^{+0.06}_{-0.06}$ & 1 \\
HD~210839       & $\lambda$~Cep & O6.5Infp       &  5.05  &  0.57  &  103.83  &   $+2.61$  &  $0.833^{+0.066}_{-0.049}$ &  $+0.038$  &  $21.48^{+0.04}_{-0.04}$ & 1 \\
HD~212791       & V408~Lac & B3ne     &  8.02  &  0.17  &  101.64  &   $-4.30$  &  $0.893^{+0.019}_{-0.016}$ &  $-0.067$  &  $21.13^{+0.12}_{-0.16}$ & 2 \\
HD~218915       & \ldots & O9.2Iab     &  7.20  &  0.30  &  108.06  &   $-6.89$  &  $2.97^{+0.33}_{-0.25}$ &  $-0.357$  &  $21.27^{+0.06}_{-0.07}$ & 1 \\
HD~219188       & \ldots & B0.5IIIn     &  7.06  &  0.13  &   83.03  &  $-50.17$  &  $2.10^{+0.25}_{-0.27}$ &  $-1.611$  &  $20.76^{+0.07}_{-0.08}$ & 1 \\
HD~220057       & \ldots & B3IV         &  6.94  &  0.23  &  112.13  &   $+0.21$  &  $0.385^{+0.004}_{-0.004}$ &  $+0.001$  &  $21.10^{+0.11}_{-0.14}$ & 1 \\
HD~224151       & V373~Cas & B0.5II-III   &  6.00  &  0.44  &  115.44  &   $-4.64$  &  $1.89^{+0.13}_{-0.10}$ &  $-0.153$  &  $21.47^{+0.04}_{-0.05}$ & 1 \\
HDE~232522      & \ldots & B1II         &  8.70  &  0.27  &  130.70  &   $-6.71$  &  $3.46^{+0.41}_{-0.44}$ &  $-0.404$  &  $21.21^{+0.04}_{-0.04}$ & 1 \\
HDE~303308      & \ldots & O4.5Vfc         &  8.17  &  0.45  &  287.59  &   $-0.61$  &  $2.17^{+0.09}_{-0.10}$ &  $-0.023$  &  $21.46^{+0.03}_{-0.03}$ & 1 \\
HDE~308813      & \ldots & O9.7IVn        &  9.73  &  0.34  &  294.79  &   $-1.61$  &  $2.43^{+0.11}_{-0.09}$ &  $-0.068$  &  $21.28^{+0.05}_{-0.06}$ & 1 \\
CPD$-$59~2603   & V572~Car & O7Vnz         &  8.81  &  0.46  &  287.59  &   $-0.69$  &  $2.63^{+0.16}_{-0.14}$ &  $-0.032$  &  $21.46^{+0.04}_{-0.04}$ & 1 \\
\enddata
\tablenotetext{a}{Distances are based on Gaia EDR3 parallax measurements \citep{bj21}.}
\tablenotetext{b}{Reference(s) for the values of $N$(H~{\sc i}) and $N$(H$_2$) used to calculate $N$(H$_{\rm tot}$): (1) \citet{j19}; (2) \citet{j09}; (3) \citet{w99}; (4) \citet{w20}; (5) \citet{ds94}; (6) \citet{r09}; (7) \citet{r23}.}
\end{deluxetable*}

After our initial survey of the literature, our database contained column density measurements for the dominant ions of the elements listed above for a total of 223 sight lines. All of these measurements (with the exception of those for Ti~{\sc ii}) are based on spectroscopic observations acquired using HST or FUSE. However, approximately half of the sight lines in this initial database had abundance measurements for only a few elements. In order for the method described in Section~\ref{sec:method} to yield reliable determinations of $[{\rm M}/{\rm H}]_{\rm ISM}$, each sight line should have a sufficient number of elemental abundance measurements, and the elements measured should have as wide a range as possible of $A_X$ values, so that the slopes and $y$-intercepts of the linear fits are adequately constrained. We decided (somewhat arbitrarily) that each sight line should have abundance measurements for at least eight different elements in order to be included in our survey. The imposition of this threshold left us with a sample of 78 sight lines. (Six sight lines with the requisite number of column density measurements were excluded from our survey because they do not have published and/or reliable values of $N({\rm H}~\textsc{i})$ and $N({\rm H}_2)$ necessary for the metallicity analysis.) Six additional sight lines were subsequently added to our sample based on new column density measurements (as described in Section~\ref{subsec:new_data}) bringing the total number of sight lines in our final sample to 84.

Basic information regarding the background stars and the characteristics of the sight lines included in the final sample is provided in Table~\ref{tab:sample}. The stellar distances and their uncertainties (as well as the values of $z$ derived from the distances) are based on Gaia EDR3 parallax measurements \citep{bj21}. Most of the measurements of $N({\rm H}~\textsc{i})$ and $N({\rm H}_2)$ used to derive $N({\rm H}_{\rm tot})$ are obtained from \citet{j19} or from the values compiled by \citet{j09}. Exceptions to this are noted in the table. A compilation of column density measurements for all of the elements other than hydrogen for the 84 sight lines in our final sample is provided in the appendix.

\startlongtable
\begin{deluxetable}{lcccc}
\tablecolumns{5}
\tablecaption{Wavelengths and Oscillator Strengths\label{tab:fvalues}}
\tablehead{ \colhead{Species} & \colhead{$\lambda$\tablenotemark{a}} & \colhead{$\log~f\lambda$} & \colhead{Error\tablenotemark{b}} & \colhead{Ref.} \\
\colhead{} & \colhead{(\AA{})} & \colhead{} & \colhead{(dex)} & \colhead{}}
\startdata
B~{\sc ii}    & 1362.463  &\phantom{+}3.133  &  0.002 & 1 \\
C~{\sc ii}    & 1334.532  &\phantom{+}2.234  & \ldots & 1 \\
              & 2325.403  & $-$3.954  &  0.003 & 1 \\
N~{\sc i}     &\phn951.079  & $-$0.795  & \ldots & 1 \\
              &\phn951.295  & $-$1.656  & \ldots & 1 \\
              &\phn953.415  &\phantom{+}1.091  & \ldots & 1 \\
              &\phn953.655  &\phantom{+}1.372  & \ldots & 1 \\
              &\phn959.494  & $-$1.304  & \ldots & 1 \\
              & 1134.165  &\phantom{+}1.219  & \ldots & 1 \\
              & 1134.415  &\phantom{+}1.512  & \ldots & 1 \\
              & 1134.980  &\phantom{+}1.674  & \ldots & 1 \\
              & 1159.817  & $-$1.938  & \ldots & 1 \\
              & 1160.937  & $-$2.496  & \ldots & 1 \\
              & 1199.550  &\phantom{+}2.199  & \ldots & 1 \\
              & 1200.223  &\phantom{+}2.018  & \ldots & 1 \\
              & 1200.710  &\phantom{+}1.715  & \ldots & 1 \\
O~{\sc i}     & 1355.598  & $-$2.805  & \ldots & 1 \\
Mg~{\sc ii}   & 1239.925  & $-$0.106  & \ldots & 1 \\
              & 1240.395  & $-$0.355  & \ldots & 1 \\
              & 2796.354  &\phantom{+}3.236  &  0.004 & 1 \\
              & 2803.532  &\phantom{+}2.933  &  0.003 & 1 \\
Si~{\sc ii}   & 1808.013  &\phantom{+}0.575  &  0.040 & 1 \\
              & 2335.123  & $-$2.003  &  0.080 & 1 \\
P~{\sc ii}    & 1152.818  &\phantom{+}2.496  &  0.044 & 2 \\
              & 1301.874  &\phantom{+}1.407  &  0.042 & 3 \\
              & 1532.533  &\phantom{+}1.053  &  0.044 & 4 \\
Cl~{\sc i}    & 1004.678  &\phantom{+}1.677  &  0.032 & 5 \\
              & 1094.769  &\phantom{+}1.625  &  0.012 & 5 \\
              & 1097.369  &\phantom{+}0.985  &  0.070 & 1 \\
              & 1347.240  &\phantom{+}2.314  &  0.030 & 1 \\
              & 1379.528  &\phantom{+}0.569  & \ldots & 6 \\
Cl~{\sc ii}   & 1071.036  &\phantom{+}1.182  &  0.021 & 7 \\
Ti~{\sc ii}   & 3072.970  &\phantom{+}2.571  &  0.027 & 1 \\
              & 3229.190  &\phantom{+}2.346  &  0.026 & 1 \\
              & 3241.983  &\phantom{+}2.876  &  0.027 & 1 \\
              & 3383.759  &\phantom{+}3.084  &  0.034 & 1 \\
Cr~{\sc ii}   & 2056.257  &\phantom{+}2.326  &  0.024 & 1 \\
              & 2062.236  &\phantom{+}2.194  &  0.023 & 1 \\
              & 2066.164  &\phantom{+}2.024  &  0.025 & 1 \\
Mn~{\sc ii}   & 1197.184  &\phantom{+}2.248  & \ldots & 8 \\
              & 1199.391  &\phantom{+}2.143  & \ldots & 8 \\
              & 1201.118  &\phantom{+}2.004  & \ldots & 8 \\
              & 2305.714  &\phantom{+}0.423  &  0.050 & 1 \\
              & 2576.877  &\phantom{+}2.969  &  0.006 & 1 \\
              & 2594.499  &\phantom{+}2.860  &  0.020 & 1 \\
              & 2606.462  &\phantom{+}2.712  &  0.021 & 1 \\
Fe~{\sc ii}   & 1055.262  &\phantom{+}0.812  & \ldots & 1 \\
              & 1112.048  &\phantom{+}0.695  & \ldots & 1 \\
              & 1121.975  &\phantom{+}1.512  & \ldots & 1 \\
              & 1125.448  &\phantom{+}1.244  & \ldots & 1 \\
              & 1127.098  &\phantom{+}0.102  & \ldots & 1 \\
              & 1133.665  &\phantom{+}0.728  & \ldots & 1 \\
              & 1142.366  &\phantom{+}0.661  & \ldots & 1 \\
              & 1143.226  &\phantom{+}1.342  & \ldots & 1 \\
              & 1144.938  &\phantom{+}1.978  &  0.030 & 1 \\
              & 1608.451  &\phantom{+}1.968  &  0.026 & 1 \\
              & 1611.201  &\phantom{+}0.347  &  0.080 & 1 \\
              & 2234.447  & $-$1.540  &  0.050 & 9 \\
              & 2249.877  &\phantom{+}0.612  &  0.030 & 1 \\
              & 2260.781  &\phantom{+}0.742  &  0.030 & 1 \\
              & 2344.214  &\phantom{+}2.427  &  0.008 & 1 \\
              & 2367.591  & $-$0.713  &  0.072 & 9 \\
              & 2374.461  &\phantom{+}1.871  &  0.020 & 1 \\
              & 2382.765  &\phantom{+}2.882  &  0.005 & 1 \\
Ni~{\sc ii}   & 1317.217  &\phantom{+}1.876  &  0.043 & 10 \\
              & 1370.132  &\phantom{+}1.906  &  0.042 & 10 \\
              & 1454.842  &\phantom{+}1.505  &  0.029 & 11 \\
              & 1709.604  &\phantom{+}1.735  &  0.024 & 11 \\
              & 1741.553  &\phantom{+}1.876  &  0.024 & 11 \\
              & 1751.916  &\phantom{+}1.691  &  0.023 & 11 \\
              & 1804.473  &\phantom{+}0.919  &  0.053 & 11 \\
Cu~{\sc ii}   & 1358.773  &\phantom{+}2.569  &  0.042 & 12 \\
Zn~{\sc ii}   & 2026.137  &\phantom{+}3.007  &  0.030 & 1 \\
              & 2062.660  &\phantom{+}2.706  &  0.030 & 1 \\
Ga~{\sc ii}   & 1414.402  &\phantom{+}3.399  &  0.020 & 1 \\
Ge~{\sc ii}   & 1237.059  &\phantom{+}3.033  &  0.053 & 13 \\
              & 1602.486  &\phantom{+}2.362  & \ldots & 14 \\
As~{\sc ii}   & 1263.770  &\phantom{+}2.515  & \ldots & 14 \\
Kr~{\sc i}    & 1164.867  &\phantom{+}2.341  &  0.004 & 14 \\
              & 1235.838  &\phantom{+}2.402  &  0.004 & 14 \\
Cd~{\sc ii}   & 2145.070  &\phantom{+}3.029  &  0.011 & 14 \\
              & 2265.715  &\phantom{+}2.749  &  0.006 & 14 \\
Sn~{\sc ii}   & 1400.440  &\phantom{+}3.158  &  0.040 & 14 \\
Pb~{\sc ii}   & 1203.616  &\phantom{+}2.956  &  0.017 & 15 \\     
              & 1433.906  &\phantom{+}2.663  &  0.044 & 15 \\
\enddata
\tablenotetext{a}{Wavelengths are specified in vaccuum except those for Ti~{\sc ii}, which are specified in air.}
\tablenotetext{b}{Uncertainty in the value of $\log~f\lambda$. Transitions with no uncertainties listed are derived from theoretical calculations.}
\tablerefs{(1) \citet{m03}, (2) \citet{f07}, (3) \citet{b18}, (4) \citet{r23}, (5) \citet{a19}, (6) \citet{oh13}, (7) \citet{s05}, (8) \citet{th05}, (9) \citet{m07}, (10) \citet{jt06}, (11) \citet{bb19}, (12) \citet{b09}, (13) \citet{h17}, (14) \citet{m00}, (15) \citet{h15}.}
\end{deluxetable}

All of the column densities in our final compilation were adjusted so that they correspond to a common set of oscillator strengths ($f$-values). An exhaustive list of the transitions used by various authors to derive the column densities included in the present investigation is provided in Table~\ref{tab:fvalues}. Most of the $f$-values adopted in the present work are from the compilations of \citet{m00,m03}. However, there have been several significant improvements in oscillator strengths in the two decades since those compilations were published. New experimental $f$-values are now available for commonly observed transitions of P~{\sc ii} \citep{f07,b18}, Cl~{\sc i} \citep{a19}, Cl~{\sc ii} \citep{s05}, Cu~{\sc ii} \citep{b09}, Ge~{\sc ii} \citep{h17}, and Pb~{\sc ii} \citep{h15}. In addition, empirically-derived $f$-values have been determined for several transitions of Fe~{\sc ii} \citep{m07} and Ni~{\sc ii} \citep{jt06,bb19}. There are no experimentally-determined $f$-values for the Mn~{\sc ii} $\lambda\lambda1197,1199,1201$ triplet. Following \citet{c17}, we adopt the theoretical results of \citet{th05} for these Mn~{\sc ii} lines. Finally, while there are new theoretical calculations for the oscillator strengths of the Ti~{\sc ii} $\lambda3072$, $\lambda3229$, $\lambda3241$, and $\lambda3383$ lines \citep{l16}, and the Zn~{\sc ii} $\lambda\lambda2026,2062$ doublet \citep{k15}, we prefer the experimental $f$-values listed in \citet{m03} for these transitions.\footnote{The adopted $f$-values often have little impact on the metallicity analysis. If all of the column densities for a certain element are derived from one transition (or a particular set of transitions) and the $f$-value of that transition is altered, the column densities will be shifted accordingly. However, the $B_X$ parameter will also be shifted in the opposite sense, leaving the values of $y$ for that element unchanged (see Equation 9).}

\begin{deluxetable*}{lll}
\tablecolumns{3}
\tablecaption{Reference Codes for Column Density Measurements\label{tab:codes}}
\tablehead{ \colhead{Code} & \colhead{Reference} & \colhead{Elements Included in the Present Study} }
\startdata
A++03       & \citet{a03}                  & O \\
C++91       & \citet{c91a}              & O, Mg, P, Cr, Mn, Fe, Ni, Cu, Zn ($\xi$~Per) \\
CSE91       & \citet{c91b}              & Ge ($\xi$~Per) \\
CMES93      & \citet{c93a}              & C ($\zeta$~Oph) \\
CFLT93      & \citet{c93b}              & As ($\zeta$~Oph) \\
C94         & \citet{c94}                      & Pb ($\zeta$~Oph) \\
C++94       & \citet{css94}               & Si ($\zeta$~Oph) \\
CMJS96      & \citet{c96}               & C ($\zeta$~Per) \\
CM97        & \citet{cm97}              & Kr \\
CMLS01      & \citet{c01}              & O \\
CML03       & \citet{c03}              & Kr \\
CLMS04      & \citet{c04}              & O \\
CLMS06      & \citet{c06}              & Mg, P, Mn, Ni, Cu, Ge \\
C++08       & \citet{c08}              & O, Kr \\
F++03       & \citet{f03}              & Ga ($\rho$~Oph~A) \\
H++93       & \citet{h93}                  & Cu, Ga, Ge, Kr (1 Sco) \\
J19         & \citet{j19}                       & O, Mg, Mn, Ge, Kr \\
JRS07       & \citet{j07}                 & N \\
JS07        & \citet{js07}                 & Fe \\
KAMM03      & \citet{k03}                 & N \\
KML06       & \citet{k06}                 & N \\
L++98       & \citet{l98}                & B ($\zeta$~Oph) \\
MJC98       & \citet{m98}                  & O ($\tau$~CMa) \\
M++07       & \citet{m07}                 & Si, Fe \\
RFSL11      & \citet{r11}                & B, O, Cu, Ga \\
RFL18       & \citet{r18}                & B, O, Ga, Ge, As, Kr, Cd, Sn, Pb \\
R*          & This work                            & C, O, Mg, Si, Ti, Cr, Mn, Fe, Ni, Cu, Ge, As, Kr, Cd, Pb \\
R**         & Ritchey (in preparation)                   & C, N, Mg, Si, Cr, Mn, Fe, Ni, Zn (X Per) \\
RBFS23        & \citet{r23}            & P, Cl \\
RB95        & \citet{rb95}                 & Cr, Zn \\
SCS92       & \citet{s92}                 & N, O, Mg, Cr, Mn, Fe, Ni, Cu, Ga, Ge, Kr ($\zeta$~Oph) \\
SS96a        & \citet{ss96a}              & Cr, Fe, Zn (HD 116852) \\
SS96b        & \citet{ss96b}              & P, Zn ($\zeta$~Oph) \\
SRF02       & \citet{srf02}                   & Fe \\
SCS94       & \citet{s94}                  & Si ($\zeta$~Oph) \\
SCGM97      & \citet{s97}                  & C ($\tau$~CMa) \\
SMC99       & \citet{s99}                  & Cd, Sn \\
SLMC04      & \citet{s04}                  & C \\
S++02       & \citet{s02}          & O, Mg, Mn, Fe, Ni, Cu (HD 192639) \\
S++03       & \citet{s03}          & N, Mg, Mn, Fe, Ni (HD 185418) \\
W++95       & \citet{w95}                  & Pb (1 Sco) \\
W++99       & \citet{w99}                  & C, N, O, Mg, Si, P, Cr, Mn, Fe, Ni, Cu, Zn, Ga, Ge, Sn (23 Ori) \\
W07         & \citet{w07}                         & O, Mg, Mn, Fe, Ni, Cu, Ge (HD 219188) \\
WC10        & \citet{wc10}              & Ti \\
WSSY20      & \citet{w20}                  & O, Mg, Si, P, Fe, Ni, Cu, Zn, Ga, Ge, Kr, Cd, Sn (HD 62542) \\
W*          & Welty (2022, private communication)  & Ge (HD 192639) \\
\enddata
\end{deluxetable*}

In Table~\ref{tab:codes}, we present the complete list of references that supplied column densities for the metallicity analysis, along with a code for each reference. These codes are used in the appendix to identify the source of each column density measurement. We also indicate for each reference in Table~\ref{tab:codes} the element or elements from that investigation whose column densities are included in the present study. If the column density measurements pertain to just a single line of sight, then the sight line is also indicated. It is important to note that, while over 40 references are listed in Table~\ref{tab:codes}, approximately two-thirds of the more than 900 column density measurements in our final sample come from just six references: \citet{c06}, \citet{js07}, \citet{r11,r18,r23}, and \citet{j19}. Another 12\% of the measurements are newly derived column densities obtained in this work (Section~\ref{subsec:new_data}). Finally, many of the column density measurements adopted for the line of sight toward HD~24534 (X~Per) are from an unpublished survey of atomic and molecular abundances in that direction (A.~M.~Ritchey, in preparation).

\subsection{New Column Density Determinations\label{subsec:new_data}}
Once our literature survey was complete, we recognized that there were deficiencies in the abundance data available, especially for the refractory elements Ni and Ti. These elements are particularly important in the metallicity analysis because they represent two of the most severely depleted elements. With large (negative) values of the depletion slope parameter $A_X$, these elements help to anchor the least-squares linear fits used to derive values of $[{\rm M}/{\rm H}]_{\rm ISM}$ (see Section~\ref{sec:metallicities}). However, fewer than half of the sight lines in our sample drawn from the literature had published Ni~{\sc ii} column densities \citep[e.g., from][]{c06}, while only about one-third had published measurements of Ti~{\sc ii} \citep[see][]{wc10}. This is despite the fact that Ni~{\sc ii} observations are available from the Mikulski Archive for Space Telescopes (MAST) for nearly every sight line in our sample (due to the prevalence of STIS E140H spectra obtained with the 1271~\AA{} central wavelength setting). Furthermore, nearly half of our sight lines have archival ground-based spectra obtained with the Ultraviolet and Visual Echelle Spectrograph (UVES) on the Very Large Telescope (VLT) that cover the relevant Ti~{\sc ii} transitions between 3200~\AA{} and 3400~\AA{}. We therefore decided to derive Ni~{\sc ii} column densities for all of the sight lines in our sample lacking this measurement for which high-resolution STIS echelle spectra are available in the MAST archive and Ti~{\sc ii} column densities for all of our sight lines with archival VLT/UVES data.

High-resolution (E140H and E230H) STIS spectra were obtained from the MAST archive for all central wavelength settings that cover the Ni~{\sc ii} $\lambda1317$, $\lambda1370$, $\lambda1454$, $\lambda1709$, $\lambda1741$, and $\lambda1751$ transitions. (These data have velocity resolutions in the range 2.1--3.7 km~s$^{-1}$.) Multiple exposures of the same star were co-added and the overlapping portions of the echelle orders were combined. Portions of the spectra surrounding the Ni~{\sc ii} absorption lines were normalized to the continuum via low-order Legendre polynomial fits. Column densities were obtained from individual Ni~{\sc ii} transitions by integrating the apparent optical depth (AOD) profiles \citep[e.g.,][]{ss91,j19}. Uncertainties in the AOD column densities were calculated by adding in quadrature the uncertainties arising from noise in the spectra, from errors in continuum placement, and from uncertainties in the adopted values of $\log f\lambda$. Final Ni~{\sc ii} column densities were obtained by taking a weighted mean of the results derived from the different transitions along a given line of sight.

Pipeline-processed VLT/UVES spectra covering the Ti~{\sc ii} $\lambda3229$, $\lambda3241$, and $\lambda3383$ transitions were downloaded from the European Southern Observatory (ESO) Science Archive Facility. (The UVES spectra have a nominal velocity resolution of $\sim$4.2 km~s$^{-1}$.) Multiple exposures of a given target were weighted according to the signal-to-noise (S/N) ratios of the spectra and co-added, after correcting the velocity scales to the reference frame of the local standard of rest (LSR). Column densities (and uncertainties in column density) were obtained using the AOD method, as discussed above for the Ni~{\sc ii} lines. When multiple Ti~{\sc ii} transitions were detected in a given direction, final column densities were derived from a weighted mean of the results obtained from the individual transitions.

Most of the Ni~{\sc ii} and Ti~{\sc ii} lines examined here do not show significant evidence of unresolved saturated absorption in the line profiles. The results obtained from different transitions of the same ion along the same line of sight generally agree within the uncertainties. For HD~79186 and HD~149757 ($\zeta$~Oph), however, there are significant discrepancies in the column densities derived from the Ti~{\sc ii} $\lambda3383$ and $\lambda3241$ lines. For these sight lines, we adopt the results from the stronger $\lambda3383$ transition, which is less influenced by noise arising from CCD response fringes in the UVES data. For HD~37061, the Ni~{\sc ii}~$\lambda1317$ line is the only Ni~{\sc ii} transition available, and the line both is extremely narrow and has a central depth approaching zero. For this sight line, and for several others with narrow absorption features, we obtained Ni~{\sc ii} column densities via multi-component Voigt profile fitting using the code ISMOD \citep{s08}.

We also obtained new column density determinations from high-resolution STIS spectra for two other refractory elements: Fe and Cr. Most published values of Fe~{\sc ii} column densities for sight lines in our sample come from a study that relied on low-resolution ($\Delta v\sim18$ km~s$^{-1}$) FUSE spectra \citep{js07}. Likewise, the only previously published Cr abundances in our database were derived in a single study that analyzed archival GHRS spectra \citep{rb95}. However, high-resolution STIS spectra covering the Fe~{\sc ii} $\lambda1611$, $\lambda2249$, and $\lambda2260$ transitions and the Cr~{\sc ii} $\lambda\lambda2056,2062,2066$ triplet are available for several sight lines without published abundances for these ions. We therefore obtained Fe~{\sc ii} and Cr~{\sc ii} column densities for these directions using our Voigt profile fitting technique. We also derived Mg~{\sc ii} and Mn~{\sc ii} column densities for several sight lines that had not been previously analyzed by \citet{c06} or \citet{j19}. These results were obtained via profile fitting of the Mg~{\sc ii} $\lambda\lambda1239,1240$ doublet and the Mn~{\sc ii} $\lambda1197$ and $\lambda1201$ lines.

Two sight lines (HD~73882 and HD~149404) initially had only three elemental abundance measurements published in the literature \citep[for N, Ti, and Fe;][]{k03,j07,js07,wc10}. However, new high-resolution STIS spectra recently became publically available for these directions.\footnote{\citet{r23} used these STIS data, combined with archival FUSE observations, to obtain P and Cl column densities for the line of sight to HD~73882.} Since these sight lines are also analyzed in \citet{dc21}, we wanted to perform an independent analysis of the metallicities in these directions. To accomplish this, we analyzed the O~{\sc i} $\lambda1355$, Mg~{\sc ii} $\lambda\lambda1239,1240$, Ni~{\sc ii} $\lambda1317$, Ni~{\sc ii} $\lambda1370$, Cu~{\sc ii} $\lambda1358$, and Ge~{\sc ii} $\lambda1237$ lines via profile fitting to obtain the total column densities of these ions along the lines of sight. Where multiple transitions were analyzed for a given ion, final column densities were again obtained by evaluating a weighted mean.

High-resolution STIS spectra are also available for the line of sight to HD~147933 ($\rho$~Oph~A), although, to our knowledge, these data have not yet been published. This is another important sight line that was included in the \citet{dc21} sample. The high total hydrogen column density and intrinsically narrow velocity distribution of the gas toward $\rho$~Oph~A, combined with the FUV and NUV coverage of the available high-resolution STIS spectra, allow us to derive column densities for many different dominant ions in this direction. In particular, we obtained column densities through profile synthesis fits to the following lines: C~{\sc ii}~$\lambda2325$, O~{\sc i}~$\lambda1355$, Mg~{\sc ii}~$\lambda\lambda1239,1240$, Si~{\sc ii}~$\lambda2335$, Mn~{\sc ii}~$\lambda2305$, Fe~{\sc ii}~$\lambda2367$, Ni~{\sc ii}~$\lambda1317$, Ge~{\sc ii}~$\lambda1237$, As~{\sc ii}~$\lambda1263$, Kr~{\sc i}~$\lambda1235$, Cd~{\sc ii}~$\lambda\lambda2145,2265$, and Pb~{\sc ii}~$\lambda1203$. These measurements represent the first reported detections of the weak C~{\sc ii}~$\lambda2325$ and Si~{\sc ii}~$\lambda2335$ lines toward $\rho$~Oph~A.

After compiling our database of elemental abundance measurements, we noted a few cases where the abundances appeared to be outliers. The O~{\sc i} and Kr~{\sc i} abundances toward HD~195455 reported by \citet{j19} are much larger than would be expected based on measurements toward sight lines with similar (low) molecular hydrogen fractions. The sight line to HD~195455 (at $l=20.3$, $b=-32.1$) probes the lower Galactic halo out to a distance 1.25~kpc below the Galactic plane. This sight line also exhibits an unusual pattern of elemental abundances in plots of $y$ versus $x$ (see Section~\ref{sec:metallicities}). Thus, to provide an independent check on the abundances in this direction, we derived new column densities through a profile fitting analysis of the O~{\sc i} $\lambda1355$, Mg~{\sc ii} $\lambda\lambda1239,1240$, Mn~{\sc ii} $\lambda1197$, Ni~{\sc ii} $\lambda1317$, Ge~{\sc ii} $\lambda1237$, and Kr~{\sc i} $\lambda1235$ lines. We also found that the Ge~{\sc ii} abundance toward HD~192639 reported by \citet{s02} was significantly larger than any other published measurement of Ge/H. A new column density determination for Ge~{\sc ii} toward HD~192639 (D.~E.~Welty, 2022, private communication) yields an abundance that is more in line with expectations.

\subsection{Properties of the Sight Lines\label{subsec:properties}}
Our final sample consists of 84 sight lines that probe a diverse array of interstellar environments throughout the local part of the Milky Way Galaxy. The distances to the stars used as background targets range from 120~pc to 4.0~kpc, while the values of $E(\bv)$ range from 0.07 to 0.70. Most of the targets probe interstellar material in the Galactic midplane. However, four stars (HD~116852, HD~121968, HD~195455, and HD~219188) sample gas in the lower halo out to a maximum $z$ distance of 3.3~kpc. Thirteen of the 84 stars in our sample were included in the surveys of translucent sight lines published by \citet{r02,r09}. Translucent sight lines are characterized by high visual extinction ($A_V\gtrsim1$~mag) and typically have high molecular hydrogen fractions. On the opposite extreme, 22 sight lines in our final sample have less than 10\% of their total hydrogen in molecular form. The diverse characteristics of the sight lines in our sample help to ensure that the distribution of metallicities we derive is fully representative of the ISM in the solar vicinity.

\begin{deluxetable*}{lccccc}
\tablecolumns{6}
\tablecaption{Depletion Parameters for 22 Elements\label{tab:elem_depl_par}}
\tablehead{ \colhead{Element} & \colhead{$\log(X/{\rm H})_{\sun}$\tablenotemark{a}} & \colhead{$A_X$} & \colhead{$B_X$} & \colhead{$z_X$} & \colhead{Reference} }
\startdata
B   &  $-9.15\pm0.04$ & $-1.471\pm0.110$ & $-0.801\pm0.047$ & 0.592 & 1 \\
C   &  $-3.54\pm0.04$ & $-0.101\pm0.229$ & $-0.193\pm0.060$ & 0.803 & 2 \\
N   &  $-4.10\pm0.11$ & $-0.000\pm0.079$ & $-0.109\pm0.111$ & 0.550 & 2 \\
O   &  $-3.24\pm0.05$ & $-0.284\pm0.042$ & $-0.153\pm0.051$ & 0.609 & 1 \\
Mg  &  $-4.38\pm0.02$ & $-0.997\pm0.039$ & $-0.800\pm0.022$ & 0.531 & 2 \\
Si  &  $-4.39\pm0.02$ & $-1.136\pm0.062$ & $-0.570\pm0.029$ & 0.305 & 2 \\
P   &  $-6.46\pm0.04$ & $-0.776\pm0.035$ & $-0.418\pm0.041$ & 0.520 & 3 \\
Cl  &  $-6.67\pm0.06$ & $-0.238\pm0.046$ & $-0.223\pm0.061$ & 0.593 & 3 \\
Ti  &  $-7.00\pm0.03$ & $-2.048\pm0.062$ & $-1.957\pm0.033$ & 0.430 & 2 \\
Cr  &  $-6.28\pm0.05$ & $-1.447\pm0.064$ & $-1.508\pm0.055$ & 0.470 & 2 \\
Mn  &  $-6.42\pm0.03$ & $-0.857\pm0.041$ & $-1.189\pm0.032$\tablenotemark{b} & 0.520 & 2 \\
Fe  &  $-4.46\pm0.03$ & $-1.285\pm0.044$ & $-1.513\pm0.033$ & 0.437 & 2 \\
Ni  &  $-5.71\pm0.03$ & $-1.490\pm0.062$ & $-1.829\pm0.035$ & 0.599 & 2 \\
Cu  &  $-7.66\pm0.06$ & $-0.710\pm0.088$ & $-1.118\pm0.063$\tablenotemark{b} & 0.711 & 2 \\
Zn  &  $-7.30\pm0.04$ & $-0.610\pm0.066$ & $-0.279\pm0.045$ & 0.555 & 2 \\
Ga  &  $-8.83\pm0.06$ & $-0.834\pm0.064$ & $-0.936\pm0.062$ & 0.607 & 1 \\
Ge  &  $-8.30\pm0.05$ & $-0.526\pm0.051$ & $-0.539\pm0.051$ & 0.609 & 1 \\
As  &  $-9.60\pm0.05$ & $-0.873\pm0.213$ & $-0.280\pm0.069$ & 0.856 & 1 \\
Kr  &  $-8.64\pm0.08$ & $-0.166\pm0.059$ & $-0.342\pm0.081$ & 0.663 & 1 \\
Cd  & $-10.19\pm0.03$\phn & $-0.028\pm0.221$ & $-0.108\pm0.050$ & 0.839 & 1 \\
Sn  &  $-9.81\pm0.04$ & $-0.517\pm0.070$ & $-0.148\pm0.044$ & 0.691 & 1 \\
Pb  &  $-9.87\pm0.04$ & $-1.077\pm0.396$ & $-0.179\pm0.064$ & 0.834 & 1 \\
\enddata
\tablenotetext{a}{Solar system abundances from \citet{l03}. These values are provided here because they were used in the evaluation of the $B_X$ parameters through Equation (2). They do not affect the determinations of $[{\rm M}/{\rm H}]_{\rm ISM}$.}
\tablenotetext{b}{The value of $B_X$ in this case has been adjusted to reflect the updated set of $f$-values adopted in the present investigation.}
\tablerefs{(1) \citet{r18}, (2) \citet{j09}, (3) \citet{r23}.}
\end{deluxetable*}

\section{DERIVATION OF RELATIVE ISM METALLICITIES\label{sec:metallicities}}
Once the total column densities have been determined for a variety of different elements along a given line of sight, it is straightforward to apply the methodology described in Section~\ref{sec:method} to derive the relative ISM metallicity $[{\rm M}/{\rm H}]_{\rm ISM}$ from a least-squares linear fit to a plot of $y$ versus $x$, where $y$ and $x$ are defined in Equations (9) and (10), respectively. The complete list of column density measurements used in our analysis of relative ISM metallicities is provided in the appendix. Values of the element-specific depletion parameters $A_X$, $B_X$, and $z_X$ are compiled in Table~\ref{tab:elem_depl_par} for the 22 elements considered in our investigation. Note that the solar system abundances \citep[from][]{l03} are included in Table~\ref{tab:elem_depl_par} only because they were used in the evaluation of the $B_X$ parameters through Equation (2) \citep[see][]{j09,r18,r23}. They do not affect the determinations of relative ISM metallicities. Also note that the values of $B_{\rm Mn}$ and $B_{\rm Cu}$ have been adjusted here to reflect the updated set of $f$-values adopted in the present work.

\subsection{Basic Results\label{subsec:results}}
\begin{figure*}
\centering
\includegraphics[width=0.9\textwidth]{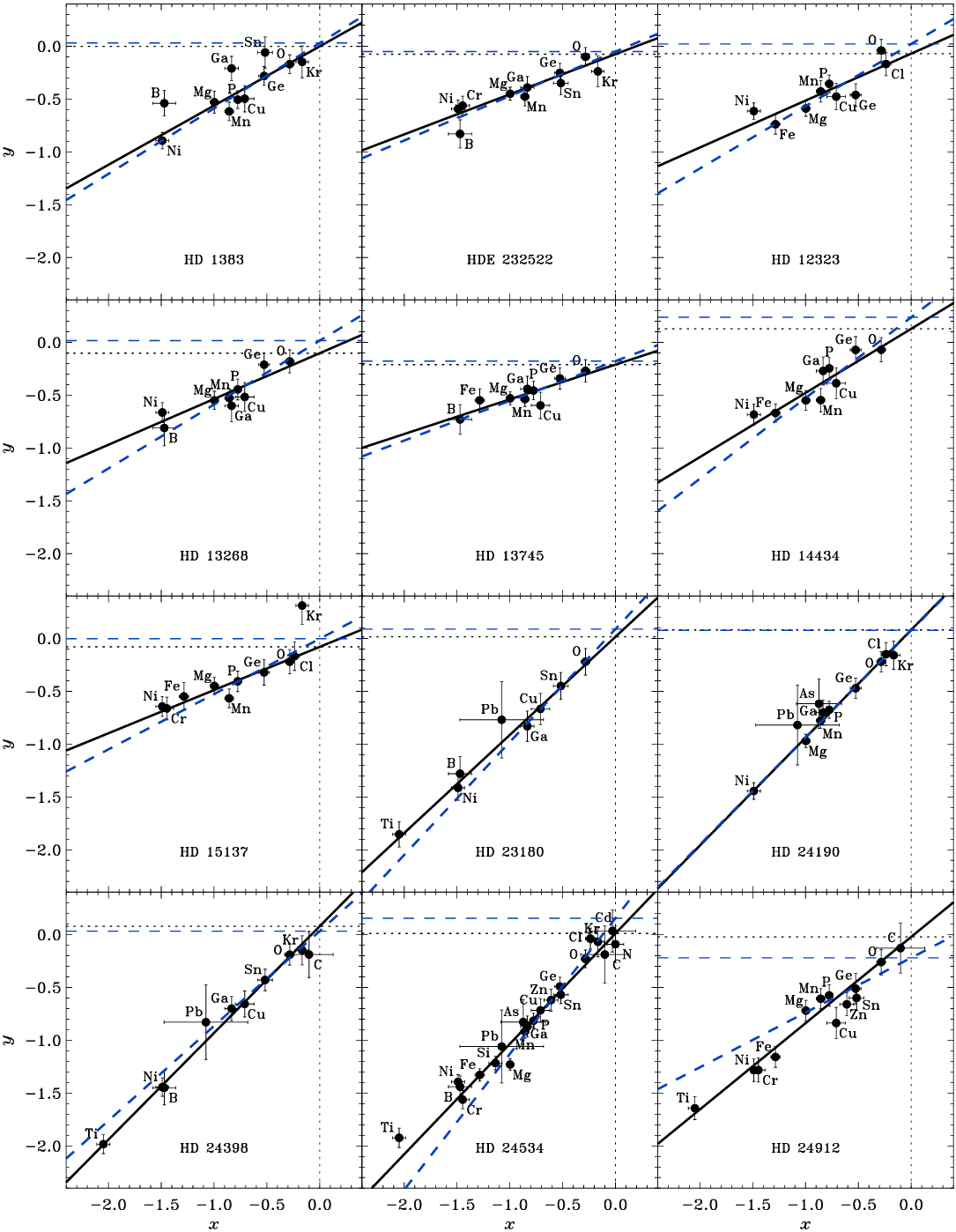}
\caption{Derivations of relative ISM metallicities. (See Equations 9 and 10 for the definitions of $y$ and $x$, respectively.) The least-squares linear fit represented by the solid black line includes measurements for all of the elements available for a given line of sight. The linear fit represented by the dashed blue line excludes the refractory elements Ti, Ni, Cr, Fe, and B. The dotted black and dashed blue horizontal lines indicate the $y$-intercepts associated with the linear fits represented by the solid black and dashed blue lines, respectively.\label{fig:metal1}}
\end{figure*}

\begin{figure*}
\centering
\includegraphics[width=0.9\textwidth]{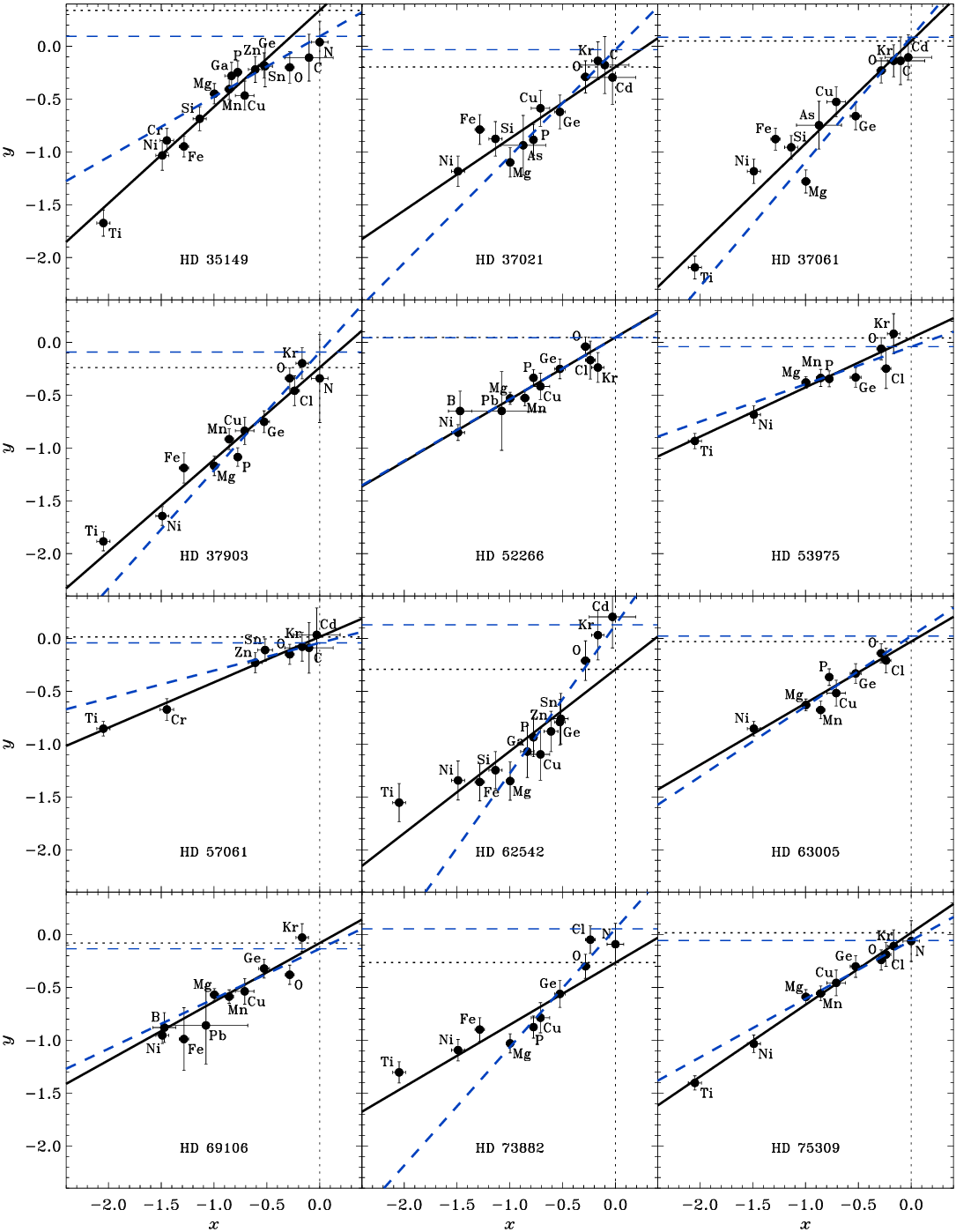}
\caption{See caption to Figure~\ref{fig:metal1}.\label{fig:metal2}}
\end{figure*}

\begin{figure*}
\centering
\includegraphics[width=0.9\textwidth]{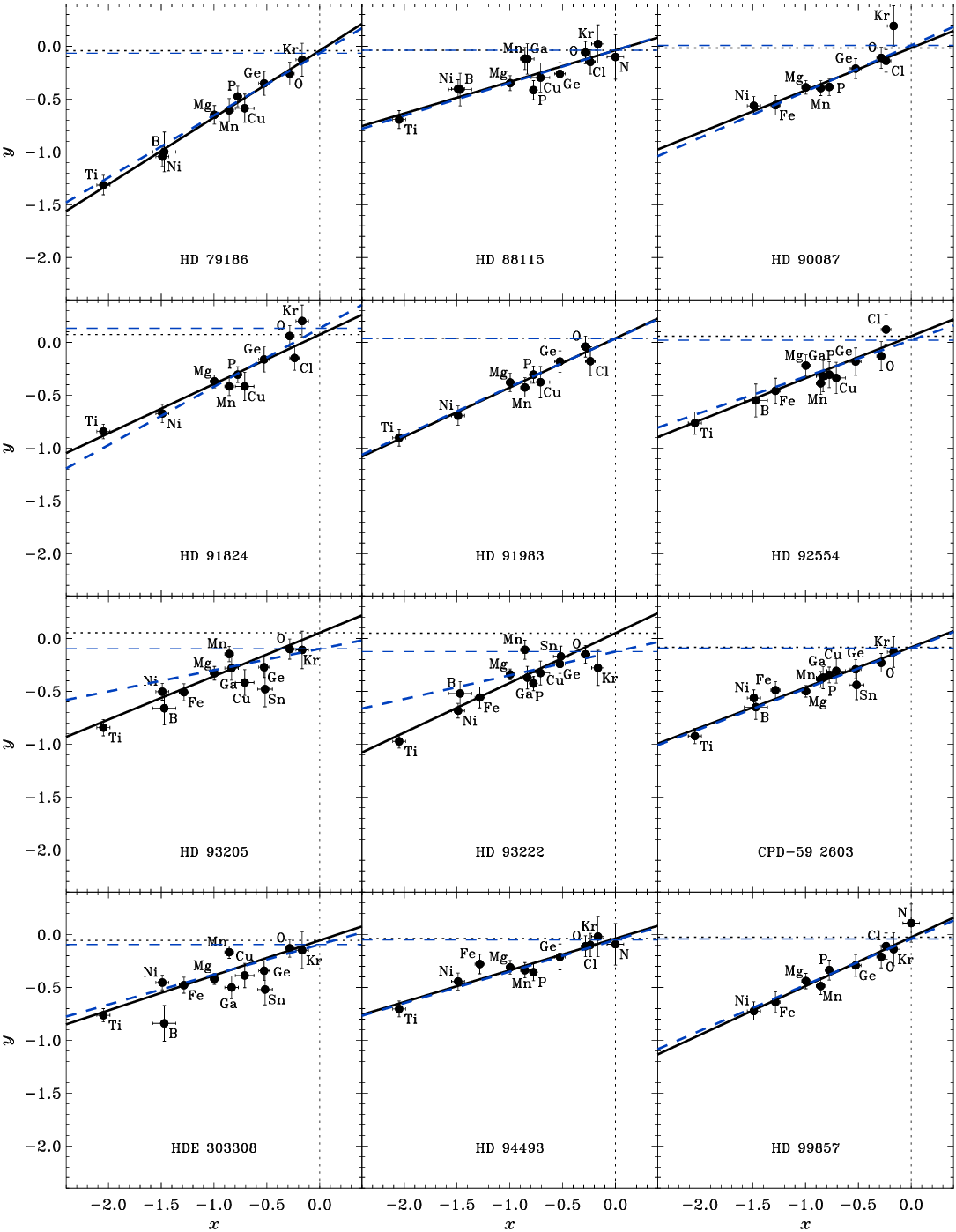}
\caption{See caption to Figure~\ref{fig:metal1}.\label{fig:metal3}}
\end{figure*}

\begin{figure*}
\centering
\includegraphics[width=0.9\textwidth]{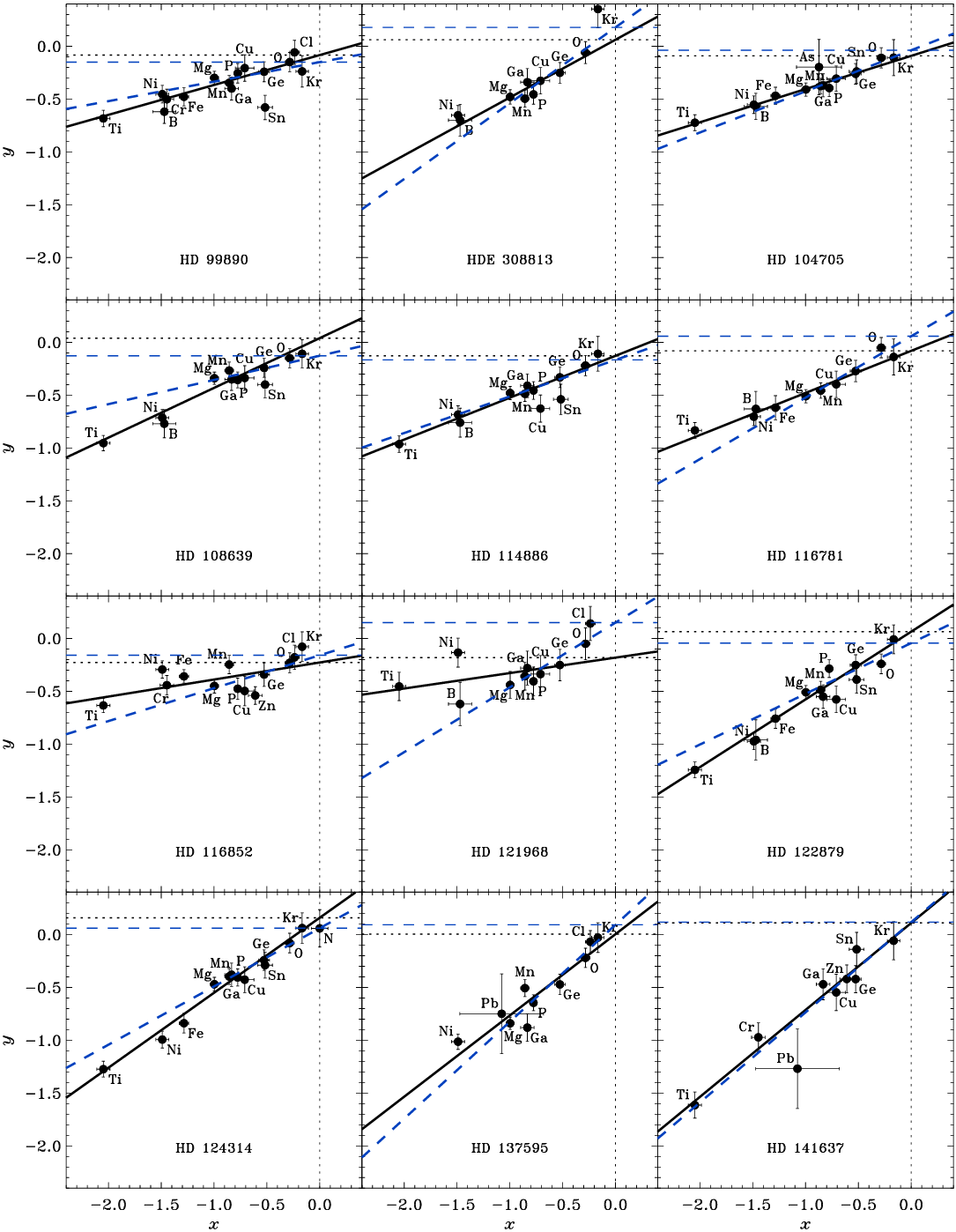}
\caption{See caption to Figure~\ref{fig:metal1}.\label{fig:metal4}}
\end{figure*}

\begin{figure*}
\centering
\includegraphics[width=0.9\textwidth]{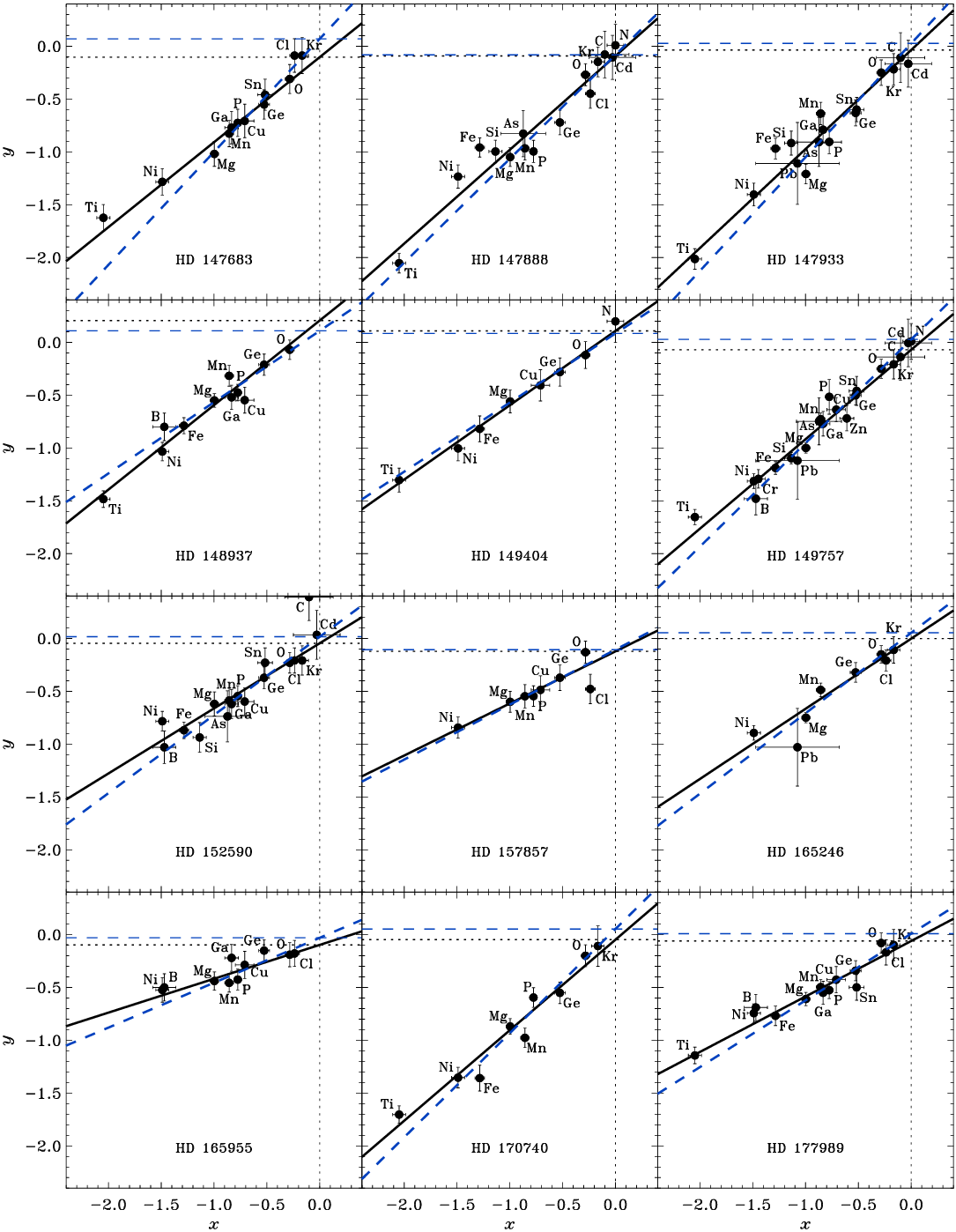}
\caption{See caption to Figure~\ref{fig:metal1}.\label{fig:metal5}}
\end{figure*}

\begin{figure*}
\centering
\includegraphics[width=0.9\textwidth]{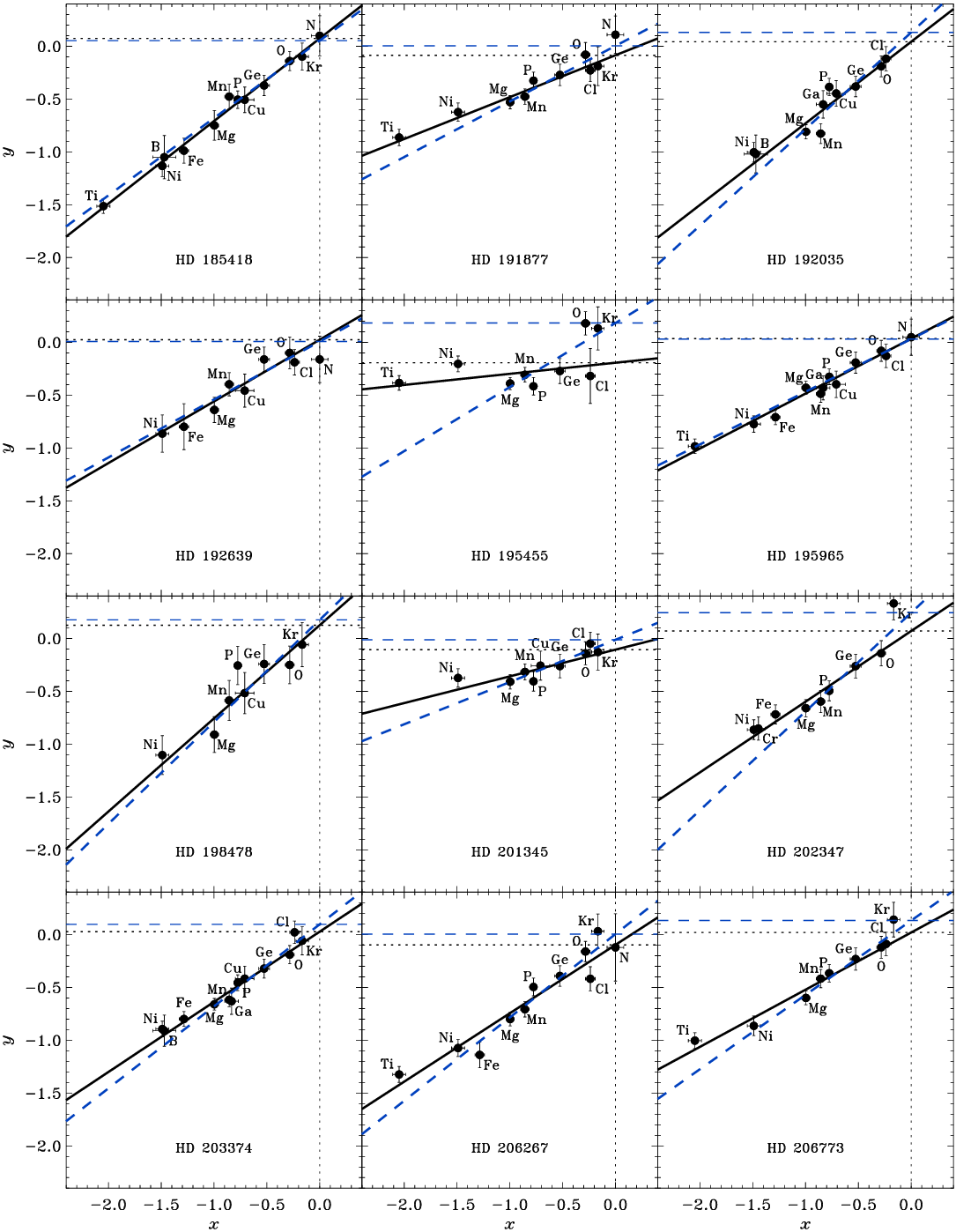}
\caption{See caption to Figure~\ref{fig:metal1}.\label{fig:metal6}}
\end{figure*}

\begin{figure*}
\centering
\includegraphics[width=0.9\textwidth]{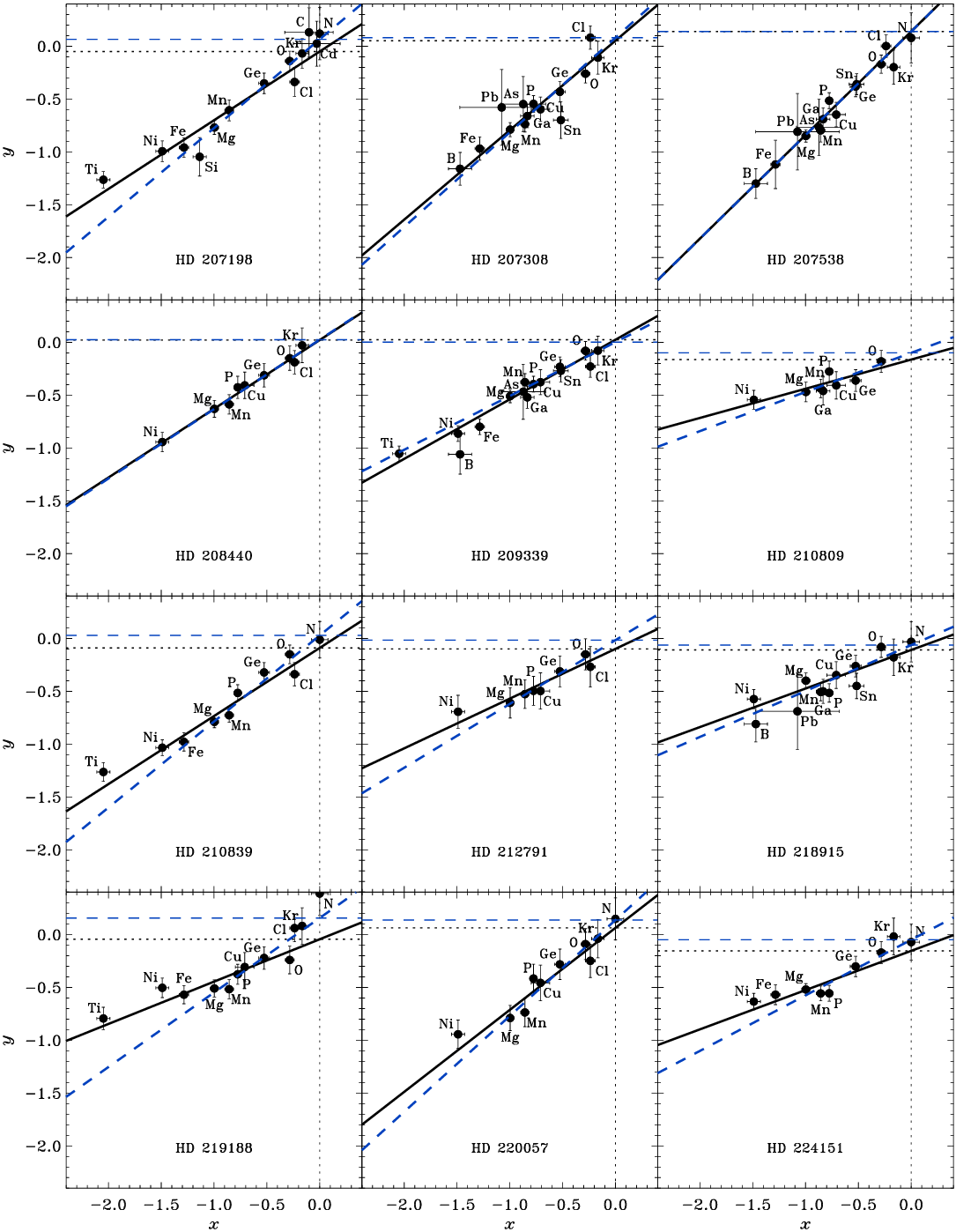}
\caption{See caption to Figure~\ref{fig:metal1}.\label{fig:metal7}}
\end{figure*}

In Figures~\ref{fig:metal1}--\ref{fig:metal7}, we present plots of $y$ versus $x$ for the 84 sight lines in our final sample. Least-squares linear fits were evaluated using the Interactive Data Language (IDL) procedure FITEXY, which accounts for errors in both the $x$ and $y$ coordinates \citep{p07}. The solid black diagonal lines in Figures~\ref{fig:metal1}--\ref{fig:metal7} represent linear fits that include all of the elemental abundance measurements available for a given sight line. The dotted black horizontal lines indicate the $y$-intercepts associated with these linear fits. The derived values for the slopes, which represent $F_*$, and the $y$-intercepts, which represent $[{\rm M}/{\rm H}]_{\rm ISM}$, along with their associated uncertainties, are presented in Table~\ref{tab:metallicities}. In each case, we provide the reduced $\chi^2$ value and the number of elemental abundance measurements included in the fit.

\subsection{Nonlinearity in Trends of $y$ versus $x$\label{subsec:nonlinear}}
Most of the linear fits that incorporate all of the abundance measurements available in a given direction are relatively good. The median value of $\chi^2/\nu$ for the fits involving all elements is 1.10. However, the results for 12 sight lines (HD~37021, HD~37061, HD~62542, HD~73882, HD~116852, HD~121968, HD~147888, HD~147933, HD~195455, HD~198478, HD~219188, and HDE~303308) indicate relatively poor fits, with $\chi^2/\nu>2$ (see Table~\ref{tab:metallicities}). In some cases, a poor fit may indicate that the errors associated with the column density determinations have been underestimated. Alternatively, there could be local enhancements in the abundances of certain elements (and not others) due to specific nucleosynthetic processes. There could also be unusual ionization conditions in certain regions that appear to enhance the abundances of dominant ions with large ionization potentials.

For several sight lines in our sample with poor least-squares linear fits, a different explanation seems more likely. For the sight lines to HD~62542, HD~73882, HD~121968, HD~195455, and HD~219188 (for example), the $y$ values that correspond to refractory elements (e.g., Fe, Ni, and Ti) are much larger than would be expected from an extrapolation of the trend seen for the more volatile elements. The most likely explanation for this behavior is that the stars in question probe multiple interstellar regions with very different depletion properties along the same line of sight. In this scenario, refractory elements will be heavily depleted relative to volatile ones in interstellar clouds with large values of $F_*$. However, these same elements will be significantly less depleted in clouds with much lower $F_*$ values. Typically, the more depleted clouds will contain the bulk of the hydrogen along the line of sight. This effect will enhance the total sight-line column densities of refractory elements since significant portions of their gas-phase abundances originate in regions with very little total hydrogen \citep[e.g., see][]{s92,ss96a,ss96b,w99}.

\startlongtable
\begin{deluxetable*}{lccccccccc}
\tablecolumns{10}
\tabletypesize{\small}
\tablecaption{Relative ISM Metallicities\label{tab:metallicities}}
\tablehead{ \colhead{Star} & \multicolumn{4}{c}{All Elements Included} & & \multicolumn{4}{c}{Refractory Elements Excluded} \\
\cline{2-5} \cline{7-10} \\
\colhead{} & \colhead{$[{\rm M}/{\rm H}]_{\rm ISM}$} & \colhead{$F_*$} & \colhead{$\chi^2/\nu$} & \colhead{$N$\tablenotemark{a}} & \colhead{} & \colhead{$[{\rm M}/{\rm H}]_{\rm ISM}$} & \colhead{$F_*$} & \colhead{$\chi^2/\nu$} & \colhead{$N$\tablenotemark{a}} }
\startdata
HD~1383        & $-0.001\pm0.077$  & $0.561\pm0.075$  & 1.96  & 11  && $+0.030\pm0.100$  & $0.619\pm0.137$  & 1.71  &  9 \\
HD~12323       & $-0.070\pm0.079$  & $0.444\pm0.077$  & 1.66  &  9  && $+0.022\pm0.096$  & $0.589\pm0.128$  & 1.23  &  7 \\
HD~13268       & $-0.102\pm0.102$  & $0.433\pm0.088$  & 0.79  &  9  && $+0.016\pm0.126$  & $0.605\pm0.149$  & 0.52  &  7 \\
HD~13745       & $-0.211\pm0.103$  & $0.328\pm0.105$  & 0.49  &  9  && $-0.177\pm0.113$  & $0.376\pm0.141$  & 0.50  &  7 \\
HD~14434       & $+0.129\pm0.104$  & $0.607\pm0.082$  & 1.40  &  9  && $+0.237\pm0.136$  & $0.764\pm0.156$  & 1.44  &  7 \\
HD~15137       & $-0.079\pm0.091$  & $0.409\pm0.066$  & 1.93  & 10  && $-0.002\pm0.099$  & $0.524\pm0.110$  & 2.69  &  7 \\
HD~23180       & $+0.016\pm0.124$  & $0.929\pm0.060$  & 0.22  &  8  && $+0.089\pm0.211$  & $1.069\pm0.253$  & 0.12  &  5 \\
HD~24190       & $+0.081\pm0.081$  & $1.020\pm0.078$  & 0.20  & 11  && $+0.077\pm0.088$  & $1.013\pm0.101$  & 0.22  & 10 \\
HD~24398       & $+0.079\pm0.087$  & $1.010\pm0.057$  & 0.13  & 10  && $+0.033\pm0.129$  & $0.897\pm0.197$  & 0.12  &  7 \\
HD~24534       & $+0.011\pm0.055$  & $1.045\pm0.050$  & 1.15  & 22  && $+0.155\pm0.068$  & $1.285\pm0.093$  & 0.47  & 17 \\
HD~24912       & $-0.021\pm0.097$  & $0.817\pm0.054$  & 1.70  & 13  && $-0.219\pm0.117$  & $0.518\pm0.132$  & 1.52  &  9 \\
HD~35149       & $+0.339\pm0.099$  & $0.913\pm0.061$  & 1.64  & 16  && $+0.094\pm0.110$  & $0.571\pm0.110$  & 0.86  & 12 \\
HD~37021       & $-0.196\pm0.142$  & $0.679\pm0.070$  & 4.97  & 12  && $-0.032\pm0.149$  & $1.009\pm0.110$  & 1.38  & 10 \\
HD~37061       & $+0.050\pm0.105$  & $0.971\pm0.054$  & 6.15  & 12  && $+0.086\pm0.119$  & $1.181\pm0.123$  & 2.82  &  9 \\
HD~37903       & $-0.238\pm0.079$  & $0.871\pm0.053$  & 1.79  & 12  && $-0.092\pm0.095$  & $1.118\pm0.125$  & 1.09  &  9 \\
HD~52266       & $+0.047\pm0.073$  & $0.588\pm0.071$  & 0.68  & 11  && $+0.044\pm0.082$  & $0.582\pm0.104$  & 0.75  &  9 \\
HD~53975       & $+0.042\pm0.062$  & $0.468\pm0.048$  & 0.74  &  9  && $-0.041\pm0.083$  & $0.355\pm0.112$  & 0.79  &  7 \\
HD~57061       & $+0.015\pm0.064$  & $0.430\pm0.046$  & 0.30  &  8  && $-0.042\pm0.102$  & $0.262\pm0.267$  & 0.21  &  6 \\
HD~62542       & $-0.292\pm0.178$  & $0.775\pm0.056$  & 5.54  & 14  && $+0.129\pm0.192$  & $1.406\pm0.135$  & 1.30  & 11 \\
HD~63005       & $-0.029\pm0.072$  & $0.584\pm0.079$  & 1.18  &  8  && $+0.024\pm0.086$  & $0.666\pm0.118$  & 1.22  &  7 \\
HD~69106       & $-0.080\pm0.080$  & $0.555\pm0.073$  & 0.99  & 10  && $-0.134\pm0.086$  & $0.473\pm0.106$  & 1.28  &  7 \\
HD~73882       & $-0.265\pm0.094$  & $0.588\pm0.050$  & 5.31  & 10  && $+0.054\pm0.108$  & $1.120\pm0.118$  & 0.75  &  7 \\
HD~75309       & $+0.017\pm0.056$  & $0.681\pm0.048$  & 0.30  & 10  && $-0.055\pm0.070$  & $0.553\pm0.107$  & 0.10  &  8 \\
HD~79186       & $-0.041\pm0.088$  & $0.633\pm0.050$  & 0.31  & 10  && $-0.066\pm0.107$  & $0.589\pm0.120$  & 0.31  &  7 \\
HD~88115       & $-0.038\pm0.073$  & $0.299\pm0.042$  & 1.33  & 13  && $-0.037\pm0.083$  & $0.310\pm0.090$  & 1.54  & 10 \\
HD~90087       & $-0.016\pm0.075$  & $0.401\pm0.068$  & 0.65  &  9  && $+0.009\pm0.081$  & $0.437\pm0.099$  & 0.77  &  7 \\
HD~91824       & $+0.074\pm0.061$  & $0.467\pm0.046$  & 1.04  & 10  && $+0.133\pm0.081$  & $0.553\pm0.109$  & 1.15  &  8 \\
HD~91983       & $+0.039\pm0.072$  & $0.467\pm0.048$  & 0.41  &  9  && $+0.035\pm0.101$  & $0.458\pm0.130$  & 0.52  &  7 \\
HD~92554       & $+0.059\pm0.103$  & $0.398\pm0.047$  & 1.44  & 11  && $+0.021\pm0.125$  & $0.345\pm0.123$  & 2.12  &  8 \\
HD~93205       & $+0.054\pm0.069$  & $0.411\pm0.045$  & 1.84  & 12  && $-0.098\pm0.089$  & $0.202\pm0.108$  & 2.08  &  8 \\
HD~93222       & $+0.051\pm0.054$  & $0.471\pm0.043$  & 1.80  & 13  && $-0.124\pm0.077$  & $0.224\pm0.103$  & 1.73  &  9 \\
HD~94493       & $-0.037\pm0.064$  & $0.300\pm0.041$  & 0.85  & 11  && $-0.051\pm0.076$  & $0.299\pm0.093$  & 0.47  &  8 \\
HD~99857       & $-0.025\pm0.078$  & $0.462\pm0.062$  & 0.51  & 10  && $-0.041\pm0.082$  & $0.435\pm0.091$  & 0.65  &  8 \\
HD~99890       & $-0.084\pm0.064$  & $0.283\pm0.039$  & 1.47  & 15  && $-0.150\pm0.077$  & $0.185\pm0.088$  & 1.95  & 10 \\
HD~104705      & $-0.091\pm0.066$  & $0.314\pm0.042$  & 0.20  & 14  && $-0.038\pm0.089$  & $0.389\pm0.105$  & 0.20  & 10 \\
HD~108639      & $+0.040\pm0.061$  & $0.471\pm0.045$  & 1.10  & 12  && $-0.127\pm0.080$  & $0.229\pm0.101$  & 0.55  &  9 \\
HD~114886      & $-0.128\pm0.068$  & $0.396\pm0.045$  & 0.78  & 12  && $-0.166\pm0.089$  & $0.346\pm0.107$  & 1.02  &  9 \\
HD~116781      & $-0.080\pm0.072$  & $0.398\pm0.047$  & 0.68  & 10  && $+0.058\pm0.091$  & $0.582\pm0.113$  & 0.29  &  6 \\
HD~116852      & $-0.227\pm0.058$  & $0.161\pm0.040$  & 2.52  & 13  && $-0.158\pm0.080$  & $0.312\pm0.109$  & 2.04  &  9 \\
HD~121968      & $-0.181\pm0.129$  & $0.146\pm0.045$  & 4.85  & 11  && $+0.150\pm0.144$  & $0.613\pm0.115$  & 0.78  &  8 \\
HD~122879      & $+0.065\pm0.069$  & $0.641\pm0.046$  & 1.23  & 13  && $-0.043\pm0.087$  & $0.480\pm0.105$  & 1.38  &  9 \\
HD~124314      & $+0.159\pm0.067$  & $0.710\pm0.046$  & 0.70  & 13  && $+0.061\pm0.080$  & $0.551\pm0.097$  & 0.26  & 10 \\
HD~137595      & $+0.004\pm0.074$  & $0.769\pm0.077$  & 1.86  & 10  && $+0.093\pm0.084$  & $0.918\pm0.105$  & 1.36  &  9 \\
HD~141637      & $+0.110\pm0.121$  & $0.823\pm0.064$  & 0.88  &  9  && $+0.118\pm0.204$  & $0.853\pm0.269$  & 0.98  &  7 \\
HD~147683      & $-0.103\pm0.120$  & $0.804\pm0.051$  & 1.24  & 12  && $+0.069\pm0.131$  & $1.069\pm0.108$  & 0.29  & 10 \\
HD~147888      & $-0.092\pm0.083$  & $0.888\pm0.048$  & 2.92  & 15  && $-0.081\pm0.092$  & $0.984\pm0.095$  & 1.18  & 12 \\
HD~147933      & $-0.036\pm0.094$  & $0.937\pm0.052$  & 3.31  & 16  && $+0.026\pm0.111$  & $1.076\pm0.111$  & 2.46  & 13 \\
HD~148937      & $+0.206\pm0.077$  & $0.800\pm0.053$  & 1.06  & 11  && $+0.109\pm0.108$  & $0.675\pm0.127$  & 1.27  &  7 \\
HD~149404      & $+0.108\pm0.114$  & $0.703\pm0.050$  & 0.31  &  8  && $+0.085\pm0.123$  & $0.654\pm0.113$  & 0.19  &  5 \\
HD~149757      & $-0.070\pm0.055$  & $0.847\pm0.048$  & 0.70  & 21  && $+0.028\pm0.068$  & $0.981\pm0.088$  & 0.44  & 16 \\
HD~152590      & $-0.045\pm0.078$  & $0.616\pm0.059$  & 0.96  & 17  && $+0.016\pm0.089$  & $0.740\pm0.102$  & 0.60  & 14 \\
HD~157857      & $-0.122\pm0.097$  & $0.493\pm0.080$  & 1.15  &  8  && $-0.106\pm0.111$  & $0.520\pm0.126$  & 1.36  &  7 \\
HD~165246      & $-0.001\pm0.064$  & $0.664\pm0.070$  & 1.34  &  8  && $+0.054\pm0.070$  & $0.761\pm0.093$  & 0.98  &  7 \\
HD~165955      & $-0.097\pm0.095$  & $0.321\pm0.083$  & 0.87  & 10  && $-0.030\pm0.105$  & $0.425\pm0.123$  & 0.93  &  8 \\
HD~170740      & $-0.048\pm0.082$  & $0.857\pm0.061$  & 1.92  &  9  && $+0.053\pm0.109$  & $0.986\pm0.146$  & 1.99  &  6 \\
HD~177989      & $-0.060\pm0.066$  & $0.525\pm0.044$  & 0.75  & 14  && $+0.009\pm0.081$  & $0.631\pm0.097$  & 0.57  & 10 \\
HD~185418      & $+0.074\pm0.067$  & $0.782\pm0.050$  & 0.20  & 12  && $+0.053\pm0.093$  & $0.733\pm0.138$  & 0.24  &  8 \\
HD~191877      & $-0.085\pm0.065$  & $0.396\pm0.045$  & 0.88  & 10  && $+0.003\pm0.076$  & $0.526\pm0.092$  & 0.67  &  8 \\
HD~192035      & $+0.043\pm0.088$  & $0.772\pm0.088$  & 1.70  & 10  && $+0.131\pm0.099$  & $0.914\pm0.128$  & 1.81  &  8 \\
HD~192639      & $+0.025\pm0.112$  & $0.584\pm0.109$  & 0.68  &  9  && $+0.008\pm0.111$  & $0.549\pm0.143$  & 0.91  &  7 \\
HD~195455      & $-0.193\pm0.061$  & $0.105\pm0.046$  & 4.68  &  9  && $+0.183\pm0.094$  & $0.606\pm0.133$  & 1.89  &  7 \\
HD~195965      & $+0.036\pm0.060$  & $0.520\pm0.044$  & 0.57  & 12  && $+0.032\pm0.078$  & $0.499\pm0.103$  & 0.46  &  9 \\
HD~198478      & $+0.126\pm0.177$  & $0.882\pm0.095$  & 3.50  &  8  && $+0.176\pm0.182$  & $0.965\pm0.126$  & 3.97  &  7 \\
HD~201345      & $-0.106\pm0.076$  & $0.253\pm0.075$  & 0.99  &  9  && $-0.012\pm0.084$  & $0.400\pm0.104$  & 0.37  &  8 \\
HD~202347      & $+0.071\pm0.098$  & $0.669\pm0.072$  & 1.76  &  9  && $+0.244\pm0.112$  & $0.935\pm0.131$  & 1.12  &  6 \\
HD~203374      & $+0.028\pm0.068$  & $0.664\pm0.062$  & 0.63  & 12  && $+0.097\pm0.076$  & $0.776\pm0.097$  & 0.51  &  9 \\
HD~206267      & $-0.097\pm0.067$  & $0.647\pm0.047$  & 1.85  & 11  && $+0.005\pm0.079$  & $0.789\pm0.099$  & 1.51  &  8 \\
HD~206773      & $+0.020\pm0.069$  & $0.541\pm0.047$  & 1.12  &  9  && $+0.133\pm0.083$  & $0.704\pm0.102$  & 0.50  &  7 \\
HD~207198      & $-0.050\pm0.067$  & $0.650\pm0.048$  & 1.11  & 13  && $+0.066\pm0.076$  & $0.840\pm0.094$  & 0.49  & 10 \\
HD~207308      & $+0.053\pm0.083$  & $0.847\pm0.089$  & 0.91  & 14  && $+0.081\pm0.088$  & $0.896\pm0.112$  & 1.02  & 12 \\
HD~207538      & $+0.140\pm0.080$  & $0.980\pm0.090$  & 0.43  & 15  && $+0.139\pm0.079$  & $0.980\pm0.101$  & 0.50  & 13 \\
HD~208440      & $+0.022\pm0.082$  & $0.651\pm0.083$  & 0.18  &  9  && $+0.025\pm0.091$  & $0.657\pm0.118$  & 0.21  &  8 \\
HD~209339      & $+0.025\pm0.059$  & $0.564\pm0.043$  & 0.75  & 15  && $+0.002\pm0.075$  & $0.509\pm0.094$  & 0.40  & 11 \\
HD~210809      & $-0.163\pm0.095$  & $0.276\pm0.084$  & 0.63  &  8  && $-0.099\pm0.123$  & $0.371\pm0.145$  & 0.62  &  7 \\
HD~210839      & $-0.090\pm0.064$  & $0.644\pm0.051$  & 1.58  & 10  && $+0.029\pm0.074$  & $0.815\pm0.098$  & 1.19  &  7 \\
HD~212791      & $-0.098\pm0.143$  & $0.471\pm0.098$  & 0.52  &  8  && $-0.015\pm0.151$  & $0.604\pm0.137$  & 0.19  &  7 \\
HD~218915      & $-0.108\pm0.083$  & $0.365\pm0.070$  & 1.55  & 13  && $-0.063\pm0.091$  & $0.435\pm0.103$  & 1.58  & 11 \\
HD~219188      & $-0.044\pm0.083$  & $0.401\pm0.049$  & 2.26  & 12  && $+0.157\pm0.095$  & $0.705\pm0.103$  & 0.94  &  9 \\
HD~220057      & $+0.064\pm0.124$  & $0.776\pm0.080$  & 1.47  & 10  && $+0.138\pm0.128$  & $0.908\pm0.104$  & 1.02  &  9 \\
HD~224151      & $-0.154\pm0.071$  & $0.372\pm0.072$  & 1.35  &  9  && $-0.049\pm0.079$  & $0.526\pm0.109$  & 1.06  &  7 \\
HDE~232522     & $-0.075\pm0.073$  & $0.380\pm0.065$  & 0.78  & 10  && $-0.050\pm0.083$  & $0.421\pm0.109$  & 0.59  &  7 \\
HDE~303308     & $-0.055\pm0.055$  & $0.331\pm0.043$  & 2.63  & 12  && $-0.093\pm0.080$  & $0.284\pm0.110$  & 3.45  &  8 \\
HDE~308813     & $+0.062\pm0.089$  & $0.547\pm0.085$  & 1.23  & 10  && $+0.178\pm0.104$  & $0.717\pm0.132$  & 1.06  &  8 \\
CPD$-$59~2603  & $-0.083\pm0.060$  & $0.380\pm0.043$  & 0.62  & 13  && $-0.092\pm0.079$  & $0.382\pm0.100$  & 0.42  &  9 \\
\enddata
\tablenotetext{a}{Number of elemental abundance measurements included in the least-squares linear fit.}
\end{deluxetable*}

A detailed abundance analysis by \citet{w20} confirms this scenario for HD~62542. Those authors derived abundances for a variety of different atomic and molecular species and reported total column densities for two separate groups of components along the line of sight. The main component near $v_{\rm LSR}=-5$~km~s$^{-1}$  is heavily depleted and is estimated to contain nearly 90\% of the total hydrogen column density. All of the other velocity components along the line of sight, which constitute very little in terms of total hydrogen, show much less depletion of refractory elements \citep[e.g., see Figure 6 in][]{w20}. If we had knowledge of the amount of hydrogen contained in individual velocity components along lines of sight showing nonlinear trends in $y$ versus $x$, then we could conceivably derive values of $[{\rm M}/{\rm H}]_{\rm ISM}$ for each component. \citet{w20} obtained estimates for the amount of hydrogen present in the two groups of components observed toward HD~62542 based, in part, on empirical correlations between $N({\rm H}~\textsc{i})$, $N({\rm H}_2)$, $N({\rm H}_{\rm tot})$ and the abundances of other interstellar constituents (e.g., CH, Na~{\sc i}, and the diffuse interstellar band at 5780~\AA{}). The issue with using a similar approach in the present investigation is that such correlations rely on implicit assumptions about the metallicities in the clouds under consideration. If the metallicity in a given parcel of gas is significantly different from the typical ISM metallicity, then the usual correlations would (presumably) no longer apply.

\begin{figure*}
\centering
\includegraphics[width=0.45\textwidth]{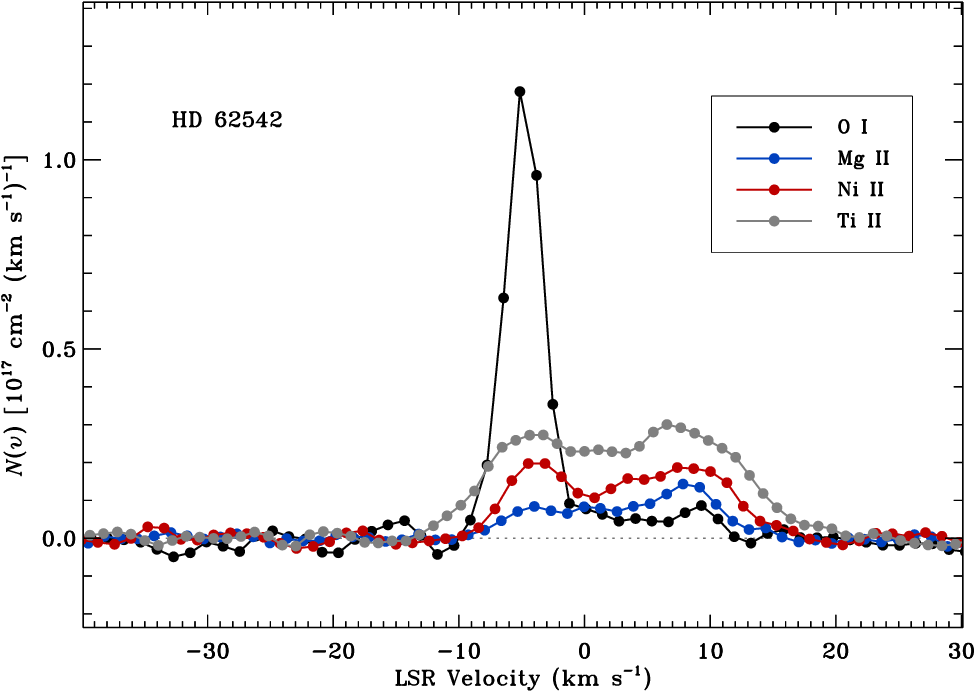}
\includegraphics[width=0.45\textwidth]{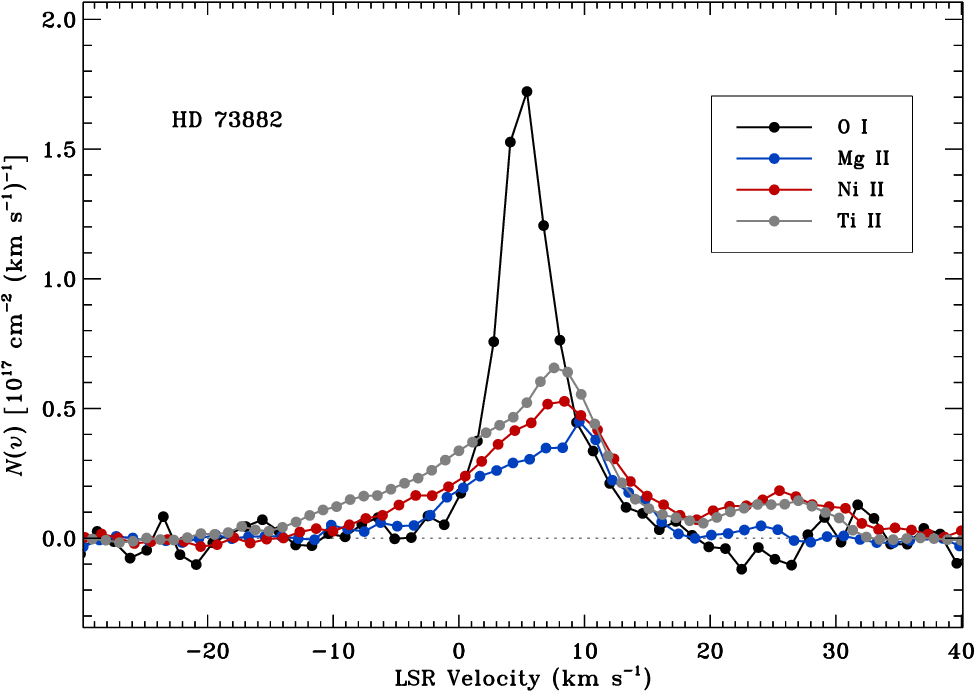}
\caption{Apparent column density profiles of O~{\sc i}, Mg~{\sc ii}, Ni~{\sc ii}, and Ti~{\sc ii} toward HD~62542 (left panel) and HD~73882 (right panel) from HST/STIS observations of the O~{\sc i}~$\lambda1355$, Mg~{\sc ii}~$\lambda1239$, and Ni~{\sc ii}~$\lambda1317$ lines and VLT/UVES observations of the Ti~{\sc ii}~$\lambda3383$ line. The column density scale applies to the O~{\sc i} profile. Profiles for the other species have been scaled to account for differences in the cosmic abundances of the elements and to adjust for the expected differences in the depletions of the elements at a particular value of $F_*$. The adopted values of $F_*$ (0.775 for HD~62542; 0.588 for HD~73882) correspond to the values obtained from our linear fits that include all of the available elements.\label{fig:profiles}}
\end{figure*}

A simple solution to the problem of nonlinearity in trends of $y$ versus $x$ is to perform a second set of least-squares linear fits in which the refractory elements are excluded. In Figures~\ref{fig:metal1}--\ref{fig:metal7}, the dashed blue diagonal lines represent linear fits that exclude the elements Ti, Ni, Cr, Fe, and B (all of which have $A_X<-1.2$). The dashed blue horizontal lines indicate the $y$-intercepts associated with these restricted fits. The resulting values of $[{\rm M}/{\rm H}]_{\rm ISM}$ and $F_*$ are again provided in Table~\ref{tab:metallicities}. With this revised set of linear fits, the median value of $\chi^2/\nu$ is now 0.92 and there are fewer cases where the reduced $\chi^2$ values are significantly larger than 2. However, the typical uncertainties in the derived values of $[{\rm M}/{\rm H}]_{\rm ISM}$ and $F_*$ are larger for these fits than for the original linear fits (because there are fewer elements considered and the elements that are included in the fits span a smaller range in $A_X$).

If we compare for a given line of sight the linear fit that includes all available elements to the linear fit restricted to non-refractory elements, we find that, in general, the restricted fits result in steeper slopes (i.e., larger sight-line depletion factors) and higher relative ISM metallicities. In most cases, the differences in the outcomes for $[{\rm M}/{\rm H}]_{\rm ISM}$ and $F_*$ between the two fits are small and within the uncertainties. Nevertheless, there does appear to be a systematic shift toward steeper slopes when the refractory elements are excluded. This indicates that the issue discussed above, where a line of sight samples multiple distinct gas regions with different depletion properties, is fairly common, although in most cases the effect appears to be rather minor.

There are four sight lines where the difference in slope between the linear fits with and without the refractory elements is larger than three times the associated uncertainty and these sight lines have been mentioned before: HD~62542, HD~73882, HD~121968, and HD~195455. Both HD~62542 and HD~73882 are well-known examples of translucent sight lines \citep[e.g.,][]{s00,r02,s07,w20}. The dominant velocity components in these directions show heavy depletions. (Our fits that exclude refractory elements indicate that $F_*>1$ in both cases.) However, both sight lines also show additional velocity components where the gas-phase abundances of refractory elements are enhanced (see Figure~\ref{fig:profiles}). The other two sight lines showing large discrepancies in the derived slope parameters are different. Both HD~121968 and HD~195455 are located at high Galactic latitude. (HD~121968 is the star that has the largest $z$ distance in Table~\ref{tab:sample}.) These two sight lines appear to probe a combination of disk gas with moderate depletions and halo gas with very low depletions. In all four cases, however, it is the sampling of multiple distinct gas regions with different depletion properties that gives rise to the nonlinear trends in plots of $y$ versus $x$.

In Figure~\ref{fig:profiles}, we provide a demonstration of the extreme differences in the velocity distributions of volatile and refractory species toward HD~62542 and HD~73882. In this figure, we plot the apparent column densities of O~{\sc i}, Mg~{\sc ii}, Ni~{\sc ii}, and Ti~{\sc ii} as a function of velocity. The Mg~{\sc ii}, Ni~{\sc ii}, and Ti~{\sc ii} profiles have been scaled to that of O~{\sc i}, accounting for differences in the cosmic abundances of the elements (Table~\ref{tab:elem_depl_par}) and adjusting for the expected differences in the depletions of the elements at a particular value of $F_*$. The adopted values of $F_*$ (0.775 for HD~62542; 0.588 for HD~73882) correspond to the values obtained from our linear fits that include all of the available elements in these directions.

Clearly, there are significant differences in the depletion properties of the various gas components seen toward HD~62542 and HD~73882. At the velocity of the dominant absorption component in each direction, the apparent column density profile of O~{\sc i} is greatly enhanced compared to those of the more refractory species. This indicates that a much larger value of $F_*$ is needed to characterize the absorption. Indeed, \citet{w20} find a value of $F_*\approx1.5$ for the main component toward HD~62542. This is similar to the value we obtain for this sight line ($F_*\approx1.4$) from our linear fit that excludes the most refractory elements.

The velocity distributions of the more refractory species toward HD~62542 and HD~73882 are much broader (and show additional peaks) compared to the relatively narrow O~{\sc i} profiles. The enhanced column densities of refractory elements indicate that much lower values of $F_*$ are required for these additional components. (\citet{w20} find that a value of $F_*\approx0.3$ characterizes the ``other'' components toward HD~62542.) However, the gas-phase abundances of Ni~{\sc ii} and Ti~{\sc ii} are enhanced (relative to Mg~{\sc ii}) even at the velocities of the dominant absorption components toward HD~62542 and HD~73882. No single value of $F_*$ can simultaneously account for the observed column densities of O~{\sc i}, Mg~{\sc ii}, Ni~{\sc ii}, and Ti~{\sc ii} in these strongly-depleted components. This could indicate that the depletion strength increases with depth into the cloud and/or that the different species have different volume distributions at a given velocity. Regardless, the above demonstration shows that the problem of nonlinearity in plots of $y$ versus $x$ can sometimes apply to individual velocity components, and not just to lines of sight as a whole.

\begin{figure}
\centering
\includegraphics[width=0.45\textwidth]{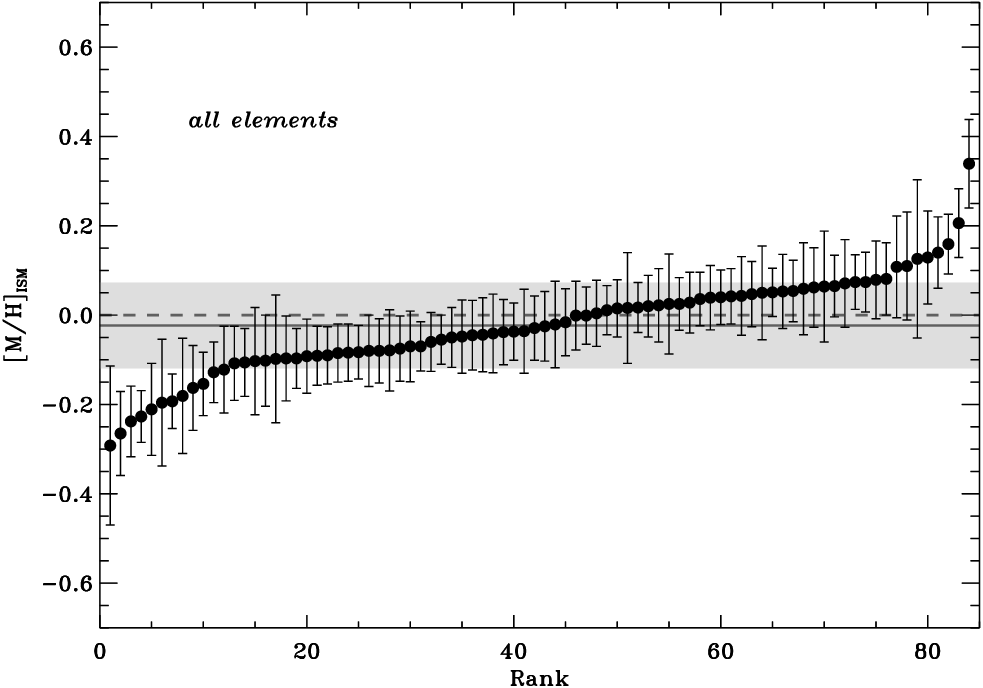}
\caption{Relative ISM metallicities determined for a sample of 84 sight lines with abundance measurements for at least eight different elements. These metallicities correspond to the linear fits that include all elements for a given line of sight. The solid dark gray horizontal line indicates the weighted mean metallicity of the sample ($-0.023\pm0.008$), while the light gray shaded region indicates the ($\pm1\sigma$) weighted standard deviation (0.096 dex). The dashed horizontal line marks the location of $[{\rm M}/{\rm H}]_{\rm ISM}=0.0$.\label{fig:rank}}
\end{figure}

\subsection{Metallicity Distributions\label{subsec:distributions}}
The main objective of our investigation is to examine for a large representative sample the distribution of metallicities seen along sight lines probing the local Galactic ISM. While the method we employ is incapable of providing us with the metallicity of the ISM relative to an adopted solar (or cosmic) abundance standard, it can tell us the degree to which the metallicity varies from one sight line to another. In Figure~\ref{fig:rank}, the relative ISM metallicities for the 84 sight lines in our final sample are plotted according to their rank. These are the metallicities derived from the least-squares linear fits that include all elemental abundance measurements available for a given line of sight. The weighted mean value of $[{\rm M}/{\rm H}]_{\rm ISM}$ for this set of measurements is $-0.023\pm0.008$. (For this calculation, we have adopted weights that correspond to the inverse squares of the measurement uncertainties.) Since the metallicities being derived are measured relative to the average ISM metallicity (Section~\ref{sec:method}), one would expect this mean value to be close to zero. \citep[It need not be exactly zero, of course, since the sample of stars examined in the present survey differs from those originally used to derive the element coefficients;][]{j09,r18,r23}. Still, the mean value given above is consistent with zero at only the $3\sigma$ level. The reason for this discrepancy may be that the linear fits that include all elements tend to underestimate the slope of the trend of $y$ versus $x$ because the refractory elements have slightly enhanced gas-phase abundances relative to volatiles when the sight line probes multiple distinct regions with different depletion properties.

\begin{figure}
\centering
\includegraphics[width=0.45\textwidth]{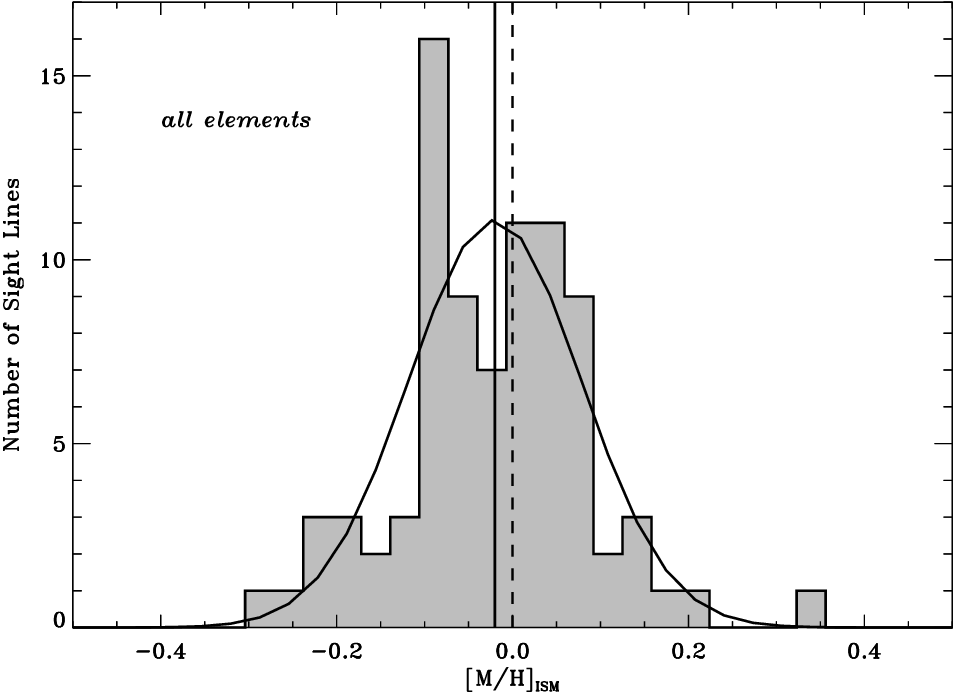}
\caption{Distribution of relative ISM metallicities. These metallicities correspond to the linear fits that include all elements for a given line of sight. A Gaussian function is fitted to the observed distribution, which is represented by a histogram. The solid vertical line indicates the mean value from the Gaussian fit ($-0.020$). The standard deviation from the fit is 0.098 dex. The dashed vertical line marks the location of $[{\rm M}/{\rm H}]_{\rm ISM}=0.0$.\label{fig:hist}}
\end{figure}

A statistic more important than the mean value in this context is the standard deviation, which should indicate the degree of chemical homogeneity exhibited by the interstellar gas in the solar neighborhood. For the linear fits that include all of the available elements, the weighted standard deviation in $[{\rm M}/{\rm H}]_{\rm ISM}$ is 0.096 dex. This may be compared to the median of the uncertainties in $[{\rm M}/{\rm H}]_{\rm ISM}$, which is 0.078 dex. To test whether the measured dispersion in $[{\rm M}/{\rm H}]_{\rm ISM}$ is significant, we can calculate a $\chi^2$ value for the sample:

\begin{equation}
\chi^2=\sum_{i=1}^N \frac{([{\rm M}/{\rm H}]_{{\rm ISM},i}-\langle [{\rm M}/{\rm H}]_{\rm ISM} \rangle)^2}{\sigma([{\rm M}/{\rm H}]_{{\rm ISM},i})^2},
\end{equation}

\noindent
where $[{\rm M}/{\rm H}]_{{\rm ISM},i}$ is the metallicity for the $i$th sight line, $\sigma([{\rm M}/{\rm H}]_{{\rm ISM},i})$ is the uncertainty in that measurement, $\langle [{\rm M}/{\rm H}]_{\rm ISM} \rangle$ is the weighted mean value derived above, and $N$ is the number of measurements. For the linear fits that include all of the available abundance measurements, the $\chi^2$ value, divided by the number of degrees of freedom, is $132.1/83=1.59$. Under the proposition that all sight lines have the same value of $[{\rm M}/{\rm H}]_{\rm ISM}$, the probability of obtaining a $\chi^2$ statistic worse than this is 0.000495. Evidently, there is a small amount of scatter that is unaccounted for by the observational uncertainties (or the observational uncertainties have been underestimated). In Figure~\ref{fig:hist}, we plot the metallicity distribution derived from the unrestricted linear fits. A Gaussian function is fitted to the observed distribution. The mean and standard deviation derived from the Gaussian fit are both nearly equal to the values obtained directly from the sample. However, the distribution appears to be somewhat irregular. In particular, there is a pileup in the number of sight lines with metallicities that are $\sim$0.09 dex below average.

\begin{figure}
\centering
\includegraphics[width=0.45\textwidth]{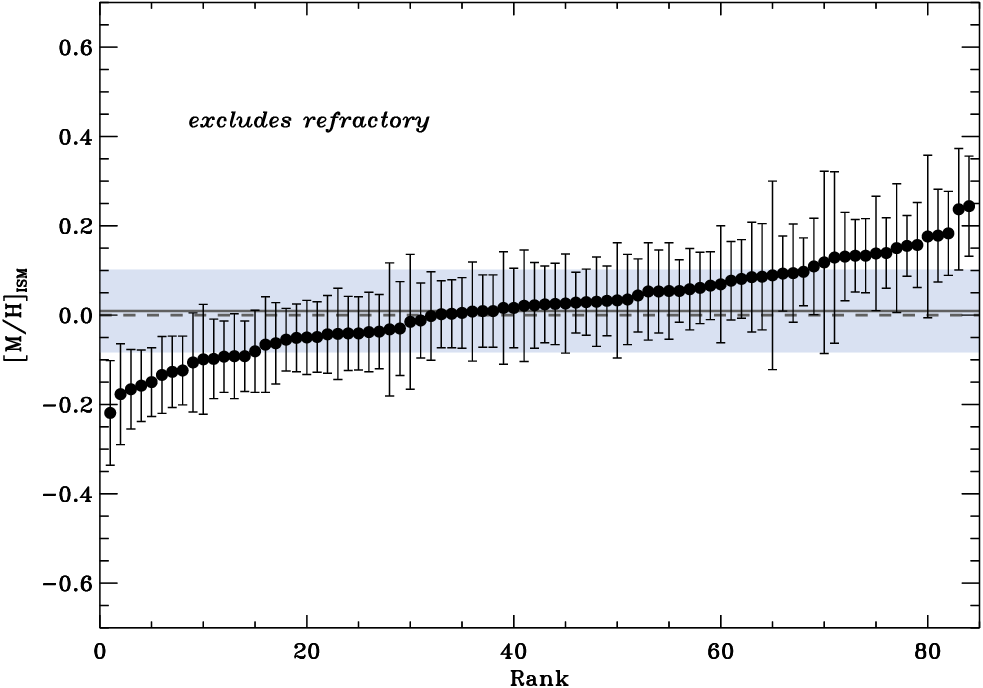}
\caption{Same as Figure~\ref{fig:rank} except that the metallicities correspond to the linear fits that exclude the refractory elements Ti, Ni, Cr, Fe, and B. The weighted mean value of $[{\rm M}/{\rm H}]_{\rm ISM}$ in this case is $+0.009\pm0.010$, while the weighted standard deviation is 0.093 dex.\label{fig:rank2}}
\end{figure}

The excess scatter in the values of $[{\rm M}/{\rm H}]_{\rm ISM}$ obtained from the unrestricted linear fits seems to arise from the varying degrees of nonlinearity observed in the trends of $y$ versus $x$ displayed in Figures~\ref{fig:metal1}--\ref{fig:metal7}. Ultimately, the cause of this variation is the extent to which different sight lines are contaminated by gas regions with depletion properties that are distinctly different from that which characterizes the bulk of the interstellar material along the line of sight. As an example, consider a sight line that probes a typical heavily-depleted diffuse molecular cloud but is contaminated by unrelated gas regions exhibiting much lower depletion strengths. In this case, the total gas-phase abundances of refractory elements will be enhanced relative to volatiles, and a linear fit that includes all elements will underpredict the slope and $y$-intercept that pertain to the dominant interstellar cloud. For a sight line such as that described here, a fit that excludes the refractory elements will provide a more accurate representation of the depletion strength and the relative ISM metallicity. (Extreme examples of this phenomenon are provided by the sight lines to HD~62542 and HD~73882; see Figures~\ref{fig:metal2} and \ref{fig:profiles}.)

\begin{figure}
\centering
\includegraphics[width=0.45\textwidth]{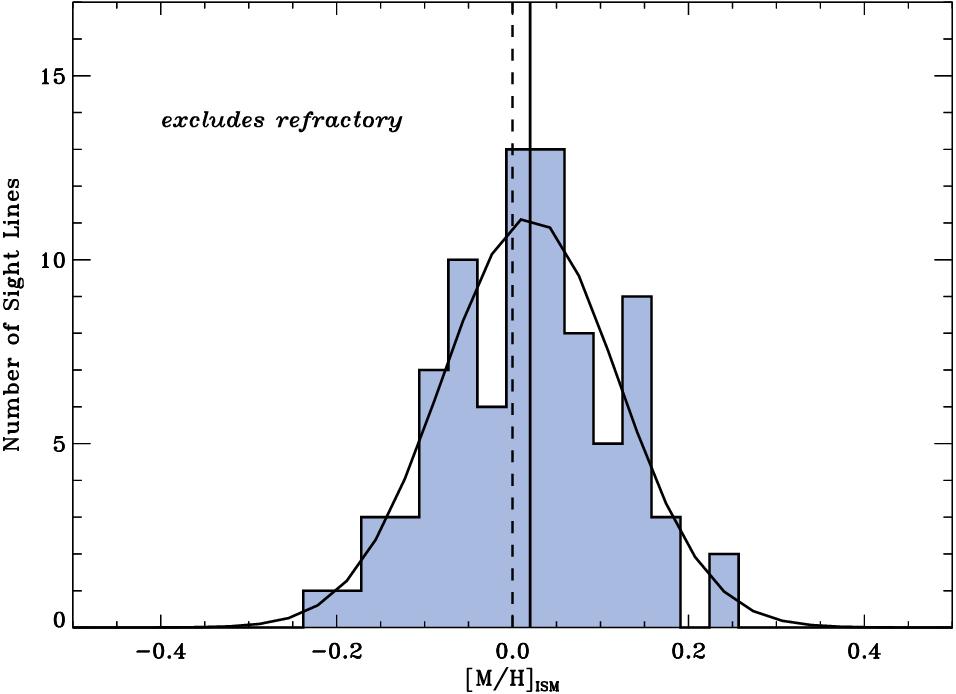}
\caption{Same as Figure~\ref{fig:hist} except that the metallicities correspond to the linear fits that exclude the refractory elements Ti, Ni, Cr, Fe, and B. The mean value from the Gaussian fit in this case is $+0.020$, while the standard deviation is 0.100 dex.\label{fig:hist2}}
\end{figure}

In Figure~\ref{fig:rank2}, we provide a plot analogous to Figure~\ref{fig:rank} showing the relative ISM metallicities derived from the linear fits that exclude the elements Ti, Ni, Cr, Fe, and B. The weighted mean value of $[{\rm M}/{\rm H}]_{\rm ISM}$ for this set of measurements is $+0.009\pm0.010$, while the weighted standard deviation is 0.093 dex. The mean value, therefore, is now statistically indistinguishable from zero and the standard deviation is somewhat smaller compared to the previous set of results. As stated earlier, the typical uncertainty in the derivations of $[{\rm M}/{\rm H}]_{\rm ISM}$ is larger for the fits restricted to non-refractory elements. Consequently, the median of the uncertainty values (0.091 dex) is now almost identical to the weighted standard deviation (0.093 dex). From Equation (15), we find that the $\chi^2/\nu$ value for this set of measurements is $83.9/83=1.01$, indicating that there is essentially no measured variation in $[{\rm M}/{\rm H}]_{\rm ISM}$ beyond that which is expected from the observational uncertainties. (The probability of obtaining a worse value for the $\chi^2$ statistic in this case is 0.451.) The analog of Figure~\ref{fig:hist} for the metallicities derived from the restricted linear fits is shown in Figure~\ref{fig:hist2}. As in the previous case, the mean and standard deviation of a Gaussian function fitted to the observed distribution are consistent with the values determined directly from the sample. However, the distribution of $[{\rm M}/{\rm H}]_{\rm ISM}$ values for this set of results is now noticably more regular.

Regardless of which set of derivations of $[{\rm M}/{\rm H}]_{\rm ISM}$ is considered, our results provide strong evidence for the chemical homogeneity of the interstellar gas in the solar neighborhood. Any metallicity variations present must be smaller than the typical measurement uncertainties. These results stand in sharp contrast to those presented by \citet{dc21}. Importantly, we see no evidence for very low metallicity gas \citep[such as that reported in][]{dc21} along any of the 84 sight lines in our sample. A detailed comparison between the metallicity derivations presented in this work and those of \citet{dc21} is provided in Section~\ref{subsec:comparison}.

\section{DISCUSSION\label{sec:discussion}}
The most significant result of our analysis of relative ISM metallicities is that the spread in metallicities exhibited by the sight lines in our sample is small and only slightly larger than the typical measurement uncertainties. Most of the column density measurements used in our metallicity analysis were derived from moderate strength atomic transitions (e.g., O~{\sc i} $\lambda1355$, Mg~{\sc ii} $\lambda\lambda1239,1240$, P~{\sc ii} $\lambda1532$, Ti~{\sc ii} $\lambda3383$, Mn~{\sc ii} $\lambda\lambda1197,1201$, Ni~{\sc ii} $\lambda1317$, Cu~{\sc ii} $\lambda1358$, Ge~{\sc ii} $\lambda1237$, and Kr~{\sc i} $\lambda1235$) recorded at high spectral resolution ($\Delta v\sim2$--4 km~s$^{-1}$). It is relatively straightforward to extract accurate column densities from such data using either the AOD or profile fitting method.\footnote{A small fraction ($\sim$11\%) of the measurements used in our analysis were derived from low-resolution FUSE observations. However, in these cases, the column densities are adequately constrained through the use of curves of growth that include very weak transitions \citep[e.g.,][]{j07,js07} or through Voigt profile fitting of weak lines, adopting component structures from higher resolution data \citep[e.g.,][]{r23}.} Furthermore, our sample is both large enough to be statistically significant and diverse enough to be representative of the local Galactic ISM. The inclusion of a wide variety of elements that exhibit a range of different depletion behaviors ensures that our linear fits are well constrained (even when the more refractory elements are excluded). It is also important to note that the elements considered in our investigation are produced through a variety of different nucleosynthetic processes. (For example, we have representatives of $\alpha$-process elements, Fe-group elements, neutron-capture elements, and elements produced through cosmic ray spallation.) This helps to ensure that we are probing variations in the overall metallicity and are not overly influenced by any potential variations in one process or another.

\subsection{Comparison with the Results of De Cia et al.\label{subsec:comparison}}
\citet{dc21} reported relative ISM metallicities for a sample of 25 sight lines probing the solar neighborhood out to a distance of 3 kpc. Most of their column density measurements were made using newly-acquired medium-resolution (E230M) STIS spectra. (STIS E230M spectra have a velocity resolution of $\sim$10~km~s$^{-1}$). \citet{dc21} derive metallicity estimates for their sight lines using two approaches. The one that they term the ``$F_*$ method'' is the same as that adopted in this investigation (see Section~\ref{sec:method}). Thus, a direct comparison of results is possible for any sight lines in common between our investigation and theirs.\footnote{It is important to understand that the metallicities derived by \citet{dc21} using the ``$F_*$ method'' are \emph{relative} ISM metallicities, just as they are in this investigation. They are not metallicities relative to an adopted solar abundance standard as is claimed in \citet{dc21}.}

There are eight sight lines in common between our metallicity study and that of \citet{dc21}: HD~23180 ($o$~Per), HD~24534 (X~Per), HD~62542, HD~73882, HD~147933 ($\rho$~Oph~A), HD~149404, HD~206267, and HD~207198. In every case, our value for the relative ISM metallicity is substantially larger than the value given by \citet{dc21}. (The differences range from 0.2 to 0.8 dex.) The largest discrepancies are found for the sight lines to X~Per, HD~62542, and HD~73882, $\rho$~Oph~A, and HD~207198. These are among the sight lines with the lowest reported metallicities in \citet{dc21} and many of these sight lines show nonlinear trends in plots of $y$ versus $x$. \citet{dc21} underestimate the metallicities in these directions primarily because they derive their estimates from only the most refractory elements, and these elements tend to exhibit shallower slopes (in the $x$-$y$ plane) compared to the more volatile elements.

The most thorough comparison between our investigation and that of \citet{dc21} can be made for the line of sight to X~Per. For this sight line, we have column density measurements for all 22 elements considered in our metallicity analysis. The uncertainties in the derived values of $[{\rm M}/{\rm H}]_{\rm ISM}$ and $F_*$ are thus among the smallest for this direction. \citet{dc21} find a depletion-corrected metallicity of $[{\rm M}/{\rm H}]=-0.57\pm0.12$ and a sight-line depletion factor of $F_*=0.61\pm0.09$ toward X~Per. However, their least-squares linear fit includes only Ti, Cr, Fe, Ni, and Zn. Abundance results for C, N, and O are plotted in the panel showing their linear fit for X~Per \citep[see Extended Data Fig.~3 in][]{dc21} but these elements are not included in the fit. The $y$ values for these volatile elements are clearly displaced upward relative to the fitted line that corresponds to the more refractory elements. Thus, if the volatile elements had been included in the fit, the slope of the fitted line would be steeper and the $y$-intercept (i.e., the metallicity) would be much larger.

From our least-squares linear fit that includes all 22 elements measured toward X~Per, we find $[{\rm M}/{\rm H}]_{\rm ISM}=+0.011\pm0.055$ and $F_*=1.045\pm0.050$ (Figure~\ref{fig:metal1}; Table~\ref{tab:metallicities}). Thus, the metallicity is indistinguishable from the average ISM metallicity, and the sight-line depletion factor is approximately equal to one, as would be expected for X~Per, which is a well-known translucent sight line. When the refractory elements are excluded from the fit, we find $[{\rm M}/{\rm H}]_{\rm ISM}=+0.155\pm0.068$ and $F_*=1.285\pm0.093$. Thus, this appears to be another case where the dominant interstellar cloud along the line of sight shows heavy depletions, but there are additional components where the gas-phase abundances of refractory elements are enhanced. (The slope parameters obtained from the two different linear fits for X~Per differ at only the $2\sigma$ level, however.)

Regardless of which of our linear fits one considers to be the most appropriate, our result for the relative ISM metallicity toward X~Per is considerably larger than that obtained by \citet{dc21}. There are two main reasons for this discrepancy. First, as already mentioned, the volatile elements (C, N, and O) are excluded from the \citet{dc21} analysis. Second, the Zn~{\sc ii} column density reported by \citet{dc21} underestimates the true column density by $\sim$0.35 dex. \citet{dc21} report a value of $\log N({\rm Zn}~\textsc{ii})=13.13\pm0.04$, which they obtain by integrating the AOD profiles of the Zn~{\sc ii} $\lambda2026$ and $\lambda2062$ lines seen in medium-resolution STIS echelle spectra.\footnote{The value quoted here for $\log N({\rm Zn}~\textsc{ii})$ from \citet{dc21} includes an adjustment of +0.10 dex to account for the difference in the adopted $f$-values of the Zn~{\sc ii} lines between our investigation and theirs. However, as mentioned in Section~\ref{subsec:lit_data}, the difference in the $f$-values does not affect the outcome for $[{\rm M}/{\rm H}]_{\rm ISM}$.} Our result of $\log N({\rm Zn}~\textsc{ii})=13.48\pm0.07$ is derived from high-resolution (E230H) STIS observations of the Zn~{\sc ii} $\lambda2062$ line (the weaker member of the Zn~{\sc ii} doublet) using a Voigt profile fitting technique. In this fit, the component structure of the relatively strong Zn~{\sc ii} $\lambda2062$ line is constrained by the results for other dominant ions with more moderate strength transitions (for more details see A.~M.~Ritchey, in preparation).

\begin{figure}
\centering
\includegraphics[width=0.45\textwidth]{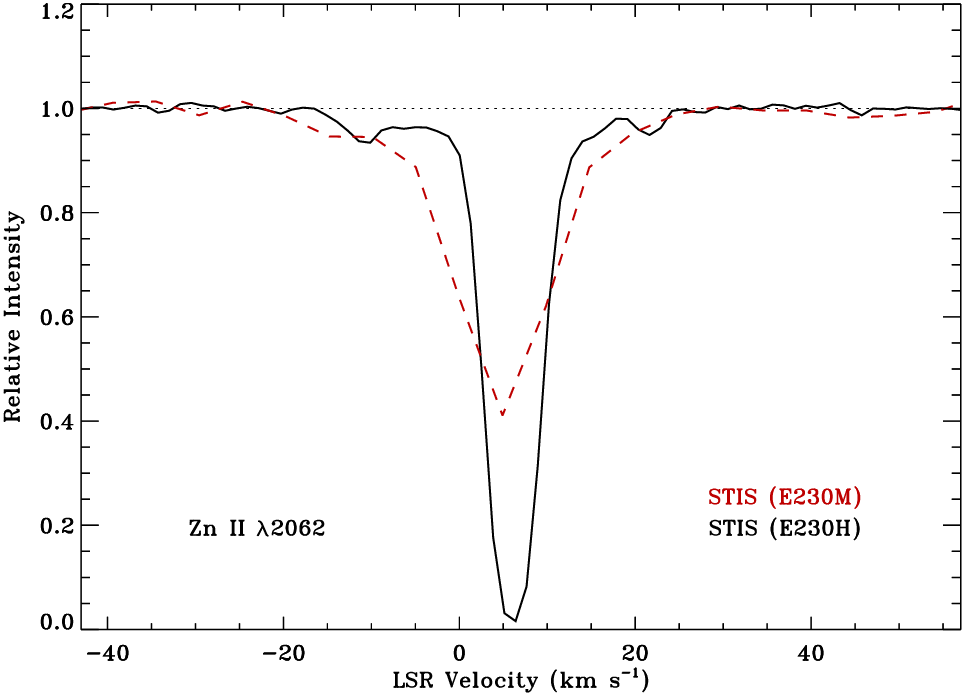}
\caption{Comparison between high-resolution (E230H) and medium-resolution (E230M) STIS spectra covering the Zn~{\sc ii} $\lambda2062$ transition toward X~Per. The Zn~{\sc ii} absorption line is heavily saturated in this direction despite the appearance of the line in the E230M spectrum.\label{fig:xper_zn}}
\end{figure}

The danger of extracting column densities from relatively strong interstellar absorption lines observed at moderate resolution using the AOD method is illustrated in Figure~\ref{fig:xper_zn}. In this figure, we compare the appearance of the Zn~{\sc ii} $\lambda2062$ line toward X~Per in medium-resolution and high-resolution STIS echelle spectra. The E230H spectrum clearly shows that the Zn~{\sc ii} line is heavily saturated despite the fact that, in the E230M spectrum, the line has a ``pointed'' appearance and the relative intensity at line center is far from zero. A straight integration of the line gives $\log N({\rm Zn}~\textsc{ii})=13.17$ for the E230H spectrum and 12.91 for E230M, while the integrated equivalent widths are nearly the same (58.5 and 58.9 m\AA{}, respectively). \citet{dc21} apply a correction to their integrated Zn~{\sc ii} column densities, based on the prescription of \citet{j96}, arriving at a value of 13.13 for X~Per. However, it appears that the optical depth correction is inadequate in this case since the integrated value from the E230H spectrum (13.17) represents a lower limit to the true column density.

The Zn~{\sc ii} column densities measured by \citet{dc21} have a significant impact on their determinations of metallicities since Zn represents one of the endpoints of their linear fits (the other endpoint being Ti). If many of their Zn~{\sc ii} column densities are underestimated, as demonstrated here for X~Per, then their metallicities are also underestimated. Indeed, by comparing the predicted values of $y$ for Zn (based on our linear fits) to the values reported by \citet{dc21} for the sight lines in common, we find that \citet{dc21} appear to have underestimated the Zn~{\sc ii} column densities by 0.1 to 0.6 dex depending on the sight line. The larger issue with the \citet{dc21} analysis, however, is that their linear fits do not include any of the relatively undepleted elements (such as C, N, O and Kr) despite the fact that measurements for these elements are readily available for many of their sight lines. Most of the volatile element measurements \citep[shown in Extended Data Fig. 3 in][]{dc21} are displaced upward relative to the fitted lines that correspond to the more refractory elements. Had these volatile elements been included in their fits, the slopes would generally be steeper and the derived metallicities would be higher.

\citet{dc21} address the discrepancy in their results between the volatile elements and the refractory ones. The same nonlinearity in plots of $y$ versus $x$ is seen for many of the sight lines in our investigation (Section~\ref{subsec:nonlinear}). However, \citet{dc21} attribute the discrepancy to a mixture of solar metallicity gas and very low metallicity (pristine) gas along the same line of sight.\footnote{These authors have softened their conclusions somewhat in an addendum to their original article \citep{dc22}.} The problem with this interpretation is that it is purely speculative. Let us consider once again the situation where a line of sight passes through a diffuse molecular cloud characterized by heavy depletions but also samples gas regions with much lower depletion strengths. According to our interpretation, the heavily depleted cloud contains the bulk of the hydrogen along the line of sight. The other components constitute very little in terms of total hydrogen yet much of the gas-phase abundances of refractory elements are contained in those minor components. \citet{dc21} would argue that those gas regions with lower depletion strengths also have much lower metallicities. However, there is no evidence to support this assertion. The problem is that we are unable to directly determine what fraction of the total hydrogen column density is associated with each of the various line-of-sight components. The \citet{dc21} interpretation implies that a large fraction of the total hydrogen column density should be attributed to those components showing weaker depletions. However, a simpler (and we would argue more likely) explanation is that the metallicities in the various line-of-sight components are roughly equivalent whereas the physical conditions (and hence the depletion strengths) are the quantities that vary from one gas region to the next.

\begin{figure}
\centering
\includegraphics[width=0.45\textwidth]{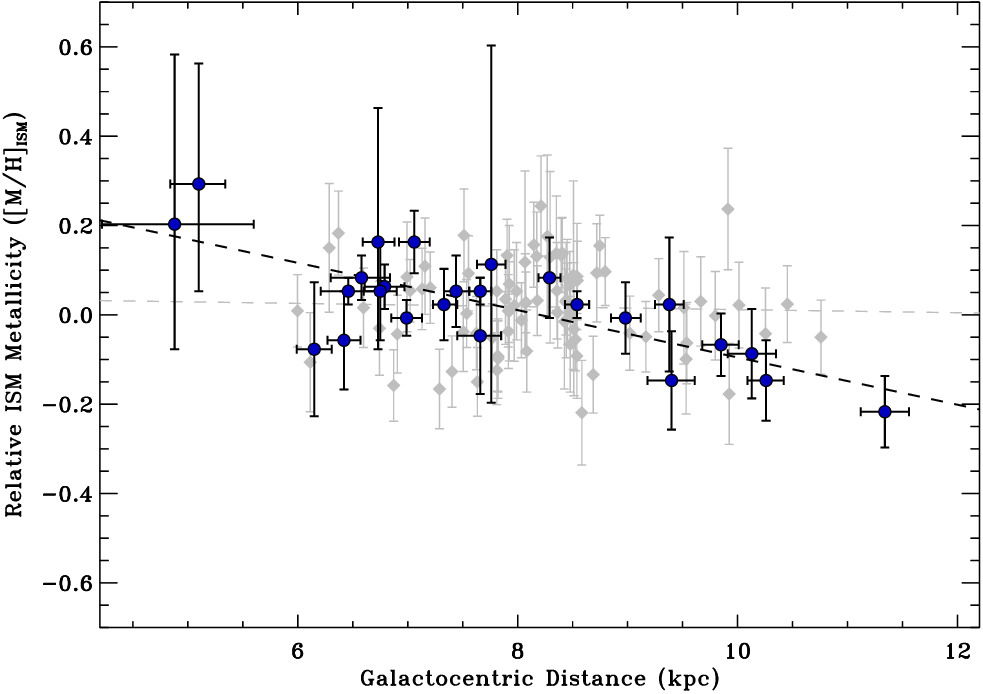}
\caption{Relative metallicities of H~{\sc ii} regions and neutral clouds in the Galactic disk plotted as a function of Galactocentric distance ($R_G$). The blue symbols represent the oxygen abundances derived for a sample of H~{\sc ii} regions located within 4 kpc of the Sun \citep{ac20,ac21}, normalized by the oxygen abundance at the solar circle (i.e., at $R_G=8.2$ kpc). The diagonal dashed line indicates the radial abundance gradient that pertains to the H~{\sc ii} regions. The light gray points represent the relative ISM metallicities derived in this work for sight lines probing neutral gas in the solar neighborhood (out to 4 kpc). The light gray dashed line indicates a linear fit to these data points. \label{fig:metal_grad}}
\end{figure}

\subsection{Metallicity Variations in the Galactic Disk\label{subsec:variations}}
Having completed our examination of the distribution of metallicities in the neutral ISM, it is instructive to compare our results to those obtained from other tracers of metallicity in the Galactic disk. Recently, \citet{e22} presented a comparison  of the metallicity distributions derived from studies of H~{\sc ii} regions, neutral clouds, B stars, Cepheids, and young clusters. The dispersions in the metallicities derived from H~{\sc ii} regions \citep{ac20,ac21}, B stars \citep{np12}, Cepheids \citep{l18}, and young clusters \citep{d20} within 3 kpc of the Sun are all similar to one another and are all $\lesssim0.10$ dex. For neutral clouds, \citet{e22} cite the results of \citet{dc21} finding that only in this case is the dispersion in metallicity much larger ($\sim$0.28 dex). If our more extensive survey of neutral cloud metallicities were considered instead, the dispersion would be $\sim$0.10 dex (Section~\ref{subsec:distributions}), similar to the values obtained from other constituents of the Galactic disk.

In Figure~\ref{fig:metal_grad}, we plot the (relative) metallicities derived for a sample of H~{\sc ii} regions located within 4 kpc of the Sun \citep{ac20,ac21} against the Galactocentric distances of those H~{\sc ii} regions \citep{md22}. The plotted values are the total oxygen abundances of the ionized nebulae normalized by the oxygen abundance at the solar circle (i.e., at $R_G=8.2$ kpc). The diagonal dashed line in the figure, which was used to determine this offset, indicates the radial abundance gradient that pertains to the H~{\sc ii} regions.\footnote{We plot \emph{relative} metallicities for the H~{\sc ii} regions in Figure~\ref{fig:metal_grad} so that the values can be more directly compared to our results for neutral clouds. Also, in this way, we avoid the so-called ``abundance discrepancy problem'', where the abundances obtained from collisionally excited lines in ionized nebulae are systematically lower than the abundances derived from recombination lines. The abundance discrepancy for O may explain why the H~{\sc ii} region O abundances are offset from the solar value by $\sim$0.2 dex \citep[e.g., see][]{ac20,e22}.} Also shown in Figure~\ref{fig:metal_grad} are the relative ISM metallicities derived in this work for sight lines probing neutral clouds in the solar neighborhood (out to 4 kpc). (For this figure, we adopt the values of $[{\rm M}/{\rm H}]_{\rm ISM}$ obtained from the fits that exclude the refractory elements.) There is no evidence of an abundance gradient from our determinations of $[{\rm M}/{\rm H}]_{\rm ISM}$, as shown by the light gray dashed line in the figure. However, the overall dispersion in the relative metallicities of the neutral clouds is similar to that of the H~{\sc ii} regions within 4~kpc of the Sun.

An important caveat regarding the comparison of relative metallicities shown in Figure~\ref{fig:metal_grad} is that, in general, the distances to the interstellar clouds seen in absorption toward background stars are not the same as the distances to the stars themselves. In contrast, the H~{\sc ii} regions are actually located at the Galactocentric distances plotted in the figure (or presumably so within the uncertainties). Typically, the dominant absorption components toward the stars in our survey are found near $v_{\rm LSR}=0$ km~s$^{-1}$. While many of the more distant stars exhibit multiple absorption complexes, corresponding to gas in different spiral arms (such as the Sagittarius-Carina spiral arm or the Perseus spiral arm), the ``local arm'' component near $v_{\rm LSR}=0$ km~s$^{-1}$ is usually the strongest. This suggests that much of the interstellar material along the various sight lines in our survey is (kinematically) close to the Sun, regardless of the distance to the background star. This may then explain why no abundance gradient is apparent in the metallicity data presented here for neutral clouds.

\section{CONCLUSIONS\label{sec:conclusions}}
In this investigation, we have presented an analysis of relative ISM metallicities for a sample of 84 sight lines probing diffuse neutral gas in the solar neighborhood out to a maximum heliocentric distance of 4 kpc. The methodology we employ was first proposed by \citet{j09} as a means to study the dust depletions and metallicities in high redshift absorption systems. The essence of this approach is to compare the pattern of relative abundances seen for a variety of elements in a given direction to the expected trends due to the depletion of atoms from the gas-phase onto interstellar dust grains. This comparison then yields the overall strength of depletion along the line of sight and whether the abundances collectively are higher or lower than the average interstellar abundances. While this method cannot provide the metallicity of the ISM with respect to some adopted cosmic abundance standard, it can yield useful information on the spread in the metallicites that pertain to neutral gas in the solar neighborhood.

To accomplish our objective of determining the metallicity distribution for a representative sample of sight lines probing the local Galactic ISM, we compiled a database of high-quality column density measurements reported in the literature for 22 elements that had previously been analyzed in accordance with the methodology devised by \citet{j09}. The vast majority of column density measurements used in the metallicity analysis were derived from observations acquired with STIS or GHRS. However, for some elements (i.e., N, Cl, and Fe), many of the column densities adopted here were obtained from lower resolution FUSE spectra. The column density measurements adopted for Ti were derived using a variety of ground-based instruments, such as VLT/UVES. We supplemented the literature survey with new column density determinations for certain key elements (e.g., Ti and Ni) and for several interesting and/or important sight lines (see Section~\ref{subsec:new_data}).

In order to properly constrain the least-squares linear fits that yielded values of $[{\rm M}/{\rm H}]_{\rm ISM}$, we required that each line of sight have column density measurements for at least eight different elements. An initial inspection of the linear fits revealed a persistent nonlinearity in the trends of $y$ versus $x$ that would cause the slope and $y$-intercept of the fit to be underestimated. We attribute this nonlinearity to a situation where a line of sight passes through multiple distinct gas regions with different depletion properties. A simple solution was to exclude the most refractory elements from the linear fits so as to obtain the depletion strength and metallicity for the bulk of the material along the line of sight.

Regardless of the set of relative ISM metallicities that is considered (Table~\ref{tab:metallicities}), the results of our analysis are clear. The dispersion in the metallicities that pertain to neutral gas in the solar neighborhood is small ($\sim$0.10 dex) and only slightly larger than the typical measurement uncertainties. These results stand in sharp contrast to those presented by \citet{dc21}, who reported metallicities as low as $-0.76$ dex, a mean metallicity of $-0.26$ dex, and a dispersion of 0.28 dex for a sample of 25 sight lines. We find no evidence for the existence of very low metallicity ($[{\rm M}/{\rm H}]\lesssim-0.3$ dex) gas along any of the 84 sight lines in our survey. There are eight sight lines in common between our investigation and that of \citet{dc21}. In each case, our value for the relative ISM metallicity is substantially larger (by 0.2 to 0.8 dex) compared to the value given in \citet{dc21}. The reason for these discrepancies is that there are serious flaws in the analysis undertaken by \citet{dc21}. The Zn~{\sc ii} column densities reported by \citet{dc21} are likely underestimated since they are based on integrated column densities derived using relatively strong absorption features recorded at moderate resolution. A more serious concern is that the metallicity determinations in \citet{dc21} are based on linear fits that include only the most refractory elements. Volatile elements, which are better tracers of metallicity in the diffuse ISM, are excluded from their analysis even though they yield conflicting results.

The dispersion in the metallicities that we obtain for sight lines probing the neutral ISM is similar to the values reported for other tracers of metallicity in the Galactic disk. The metallicity distributions for neutral clouds, H~{\sc ii} regions, B stars, Cepheids, and young clusters in the solar neighborhood all have dispersions of $\sim$0.10 dex or lower. Together, these results provide strong observational support for a well-mixed, chemically homogeneous ISM in the vicinity of the Sun.

\acknowledgments
We note the passing of our scientific colleague, Prof. Blair Savage, on July 19, 2022 in Madison, WI. Blair was a major figure in UV space astronomy who, with his students, postdocs, and colleagues, made seminal contributions to studies of gas and dust in the interstellar and intergalactic medium. We thank Dan Welty for providing useful comments on an early draft of this paper. We also thank the anonymous referee for helpful suggestions that improved our analysis. This research has made use of the SIMBAD database operated at CDS, Strasbourg, France. Observations were obtained from the ESO Science Archive Facility and the MAST data archive at the Space Telescope Science Institute. STScI is operated by the Association of Universities for Research in Astronomy, Inc., under NASA contract NAS5-26555.

\facilities{HST(STIS, GHRS), FUSE, VLT(UVES)}
\software{ISMOD \citep{s08}, STSDAS}

\appendix
\section{COMPILATION OF COLUMN DENSITY MEASUREMENTS FOR METAL IONS}
A complete compilation of the column densities used in the metallicity analysis (described in Section~\ref{sec:metallicities}) for elements other than hydrogen is provided in Table~\ref{tab:col_den}. For each entry in the table, we provide the reference for the column density measurement (see codes listed in Table~\ref{tab:codes}), the instrument that recorded the spectrum, and the method used in the analysis. The instrument codes include FUSE, GHRS, OPT, STIS, and UVES (where OPT refers to any ground-based optical telescope other than VLT/UVES). The method codes include AOD, COG, FIT, LDW, and WLL (which are explained in a footnote to Table~\ref{tab:col_den}). Note that in some cases a combination of instruments and/or methods were used. All column densities listed in Table~\ref{tab:col_den} have been adjusted so as to be consistent with the set of $f$-values provided in Table~\ref{tab:fvalues}.

In Table~\ref{tab:new_col_den}, we provide additional details regarding the column density measurements newly-derived in this investigation (see Section~\ref{subsec:new_data}). In particular, we provide the equivalent widths ($W_{\lambda}$) and column densities of the individual transitions analyzed in each direction. All of the results listed in Table~\ref{tab:new_col_den} (except those for Ti~{\sc ii}) were derived from high-resolution (E140H or E230H) STIS spectra using either the AOD or profile fitting method. The Ti~{\sc ii} column densities were extracted from archival VLT/UVES spectra using the AOD approach. In cases where multiple transitions from the same species were analyzed, final column densities were obtained by taking the weighted mean of the individual results. The final column densities in these cases are provided in Table~\ref{tab:col_den}.

\begin{deluxetable*}{lccccc}
\tablecolumns{6}
\tablecaption{Column Densities Used in the Metallicity Analysis\label{tab:col_den}}
\tablehead{ \colhead{Star} & \colhead{Species} & \colhead{$\log N$} & \colhead{Reference\tablenotemark{a}} & \colhead{Instrument} & \colhead{Method\tablenotemark{b}} }
\startdata
HD~1383       & B~{\sc ii}        & $11.92\pm0.07$   & RFSL11        & STIS        & FIT \\
       & O~{\sc i}         & $18.15\pm0.02$   & RFL18         & STIS        & FIT \\
       & Mg~{\sc ii}       & $16.36\pm0.09$   & CLMS06        & STIS        & FIT \\
       & P~{\sc ii}        & $14.56\pm0.04$   & RBFS23          & STIS        & FIT \\
       & Mn~{\sc ii}       & $13.76\pm0.06$   & CLMS06        & STIS        & FIT \\
       & Ni~{\sc ii}       & $14.00\pm0.03$   & CLMS06        & STIS        & FIT \\
       & Cu~{\sc ii}       & $12.77\pm0.05$   & RFSL11        & STIS        & FIT \\
       & Ga~{\sc ii}       & $12.07\pm0.05$   & RFSL11        & STIS        & FIT \\
       & Ge~{\sc ii}       & $12.74\pm0.02$   & RFL18         & STIS        & FIT \\
       & Kr~{\sc i}        & $12.52\pm0.08$   & RFL18         & STIS        & FIT \\
       & Sn~{\sc ii}       & $11.88\pm0.12$   & RFL18         & STIS        & FIT \\
HD~12323      & O~{\sc i}         & $18.02\pm0.06$   & CLMS04        & STIS        & FIT \\
      & Mg~{\sc ii}       & $16.04\pm0.05$   & CLMS06        & STIS        & FIT \\
      & P~{\sc ii}        & $14.45\pm0.04$   & RBFS23          & STIS        & FIT \\
      & Cl~{\sc i}+Cl~{\sc ii}   & $14.36\pm0.05$   & RBFS23          & FUSE+STIS   & FIT \\
      & Mn~{\sc ii}       & $13.69\pm0.08$   & CLMS06        & STIS        & FIT \\
      & Fe~{\sc ii}       & $15.13\pm0.07$   & JS07          & FUSE        & COG \\
      & Ni~{\sc ii}       & $14.02\pm0.03$   & CLMS06        & STIS        & FIT \\
      & Cu~{\sc ii}       & $12.53\pm0.05$   & CLMS06        & STIS        & FIT \\
      & Ge~{\sc ii}       & $12.30\pm0.06$   & CLMS06        & STIS        & FIT \\
\enddata
\tablecomments{This table is available in its entirety in machine-readable form.}
\tablenotetext{a}{Reference codes used in this table are explained in Table~\ref{tab:codes}.}
\tablenotetext{b}{Method used in the column density determination. The codes listed have the following meaning: AOD: apparent optical depth integration; COG: curve of growth analysis; FIT: Voigt profile fitting analysis; LDW: analysis of Lorentzian damping wings; WLL: weak line limit calculation.}
\end{deluxetable*}

\begin{deluxetable*}{llccc}
\tablecolumns{5}
\tablecaption{New Column Density Determinations\label{tab:new_col_den}}
\tablehead{ \colhead{Star} & \colhead{Transition} & \colhead{$W_{\lambda}$ (m\AA{})} & \colhead{$\log N$} & \colhead{Method\tablenotemark{a}} }
\startdata
HD~14434      & Fe~{\sc ii} $\lambda2249$  &  $155.4\pm3.1$\phn  &  $15.39\pm0.01$  &  FIT \\
      & Fe~{\sc ii} $\lambda2260$  &  $193.9\pm3.1$\phn  &  $15.39\pm0.01$  &  FIT \\
HD~15137      & Cr~{\sc ii} $\lambda2056$  &  $99.5\pm3.4$  &  $13.54\pm0.02$  &  FIT \\
      & Cr~{\sc ii} $\lambda2062$  &  $78.5\pm3.3$  &  $13.55\pm0.02$  &  FIT \\
      & Cr~{\sc ii} $\lambda2066$  &  $57.8\pm3.0$  &  $13.55\pm0.02$  &  FIT \\
      & Ni~{\sc ii} $\lambda1317$  &  $65.4\pm1.8$  &  $14.01\pm0.04$  &  AOD \\
      & Ni~{\sc ii} $\lambda1454$  &  $39.1\pm2.7$  &  $14.04\pm0.12$  &  AOD \\
      & Ni~{\sc ii} $\lambda1709$  &  $71.0\pm6.1$  &  $14.03\pm0.05$  &  AOD \\
      & Ni~{\sc ii} $\lambda1741$  &  $93.6\pm4.4$  &  $14.04\pm0.04$  &  AOD \\
      & Ni~{\sc ii} $\lambda1751$  &  $69.5\pm4.8$  &  $14.04\pm0.05$  &  AOD \\
HD~23180      & Ti~{\sc ii} $\lambda3383$  &  \phn$6.6\pm0.5$  &  $11.26\pm0.04$  &  AOD \\
      & Ni~{\sc ii} $\lambda1317$  &  \phn$9.9\pm0.4$  &  $13.11\pm0.02$  &  FIT \\
      & Ni~{\sc ii} $\lambda1370$  &  $12.1\pm0.3$  &  $13.15\pm0.02$  &  FIT \\
HD~24190      & Mg~{\sc ii} $\lambda1239$  &  $28.0\pm0.4$  &  $15.67\pm0.02$  &  FIT \\
      & Mg~{\sc ii} $\lambda1240$  &  $18.6\pm0.3$  &  $15.68\pm0.02$  &  FIT \\
      & Mn~{\sc ii} $\lambda1197$  &  $27.6\pm0.7$  &  $13.34\pm0.03$  &  FIT \\
      & Mn~{\sc ii} $\lambda1201$  &  $20.4\pm0.8$  &  $13.39\pm0.02$  &  FIT \\
      & Ni~{\sc ii} $\lambda1317$  &  $12.3\pm0.3$  &  $13.21\pm0.01$  &  FIT \\
      & Ni~{\sc ii} $\lambda1370$  &  $14.0\pm1.1$  &  $13.23\pm0.03$  &  FIT \\
\enddata
\tablecomments{This table is available in its entirety in machine-readable form.}
\tablenotetext{a}{Method codes used here have the same meaning as in Table~\ref{tab:col_den}.}
\end{deluxetable*}


\begin{thebibliography}{}
\bibitem[Alkhayat et al.(2019)]{a19} Alkhayat, R.~B., Irving, R.~E., Federman, S.~R., Ellis, D.~G., \& Cheng, S.\ 2019, \apj, 887, 14
\bibitem[Andr{\'e} et al.(2003)]{a03} Andr{\'e}, M.~K., Oliveira, C.~M., Howk, J.~C., et al.\ 2003, \apj, 591, 1000
\bibitem[Arellano-C{\'o}rdova et al.(2020)]{ac20} Arellano-C{\'o}rdova, K.~Z., Esteban, C., Garc{\'i}a-Rojas, J., \& M{\'e}ndez-Delgado, J.~E.\ 2020, \mnras, 496, 1051
\bibitem[Arellano-C{\'o}rdova et al.(2021)]{ac21} Arellano-C{\'o}rdova, K.~Z., Esteban, C., Garc{\'i}a-Rojas, J., \& M{\'e}ndez-Delgado, J.~E.\ 2021, \mnras, 502, 225
\bibitem[Asplund et al.(2021)]{a21} Asplund, M., Amarsi, A.~M., \& Grevesse, N.\ 2021, \aap, 653, A141
\bibitem[Bailer-Jones et al.(2021)]{bj21} Bailer-Jones, C.~A.~L., Rybizki, J., Fouesneau, M., Demleitner, M., \& Andrae, R.\ 2021, \aj, 161, 147
\bibitem[Boiss{\'e} \& Bergeron(2019)]{bb19} Boiss{\'e}, P., \& Bergeron, J.\ 2019, \aap, 622, A140
\bibitem[Brown et al.(2018)]{b18} Brown, M.~S., Alkhayat, R.~B., Irving, R.~E., et al.\ 2018, \apj, 868, 42
\bibitem[Brown et al.(2009)]{b09} Brown, M.~S., Federman, S.~R., Irving, R.~E., Cheng, S., \& Curtis, L.~J.\ 2009, \apj, 702, 880
\bibitem[Cardelli(1994)]{c94} Cardelli, J.~A.\ 1994, Sci, 265, 209
\bibitem[Cardelli et al.(1993b)]{c93b} Cardelli, J.~A., Federman, S.~R., Lambert, D.~L., \& Theodosiou, C.~E.\ 1993b, \apjl, 416, L41
\bibitem[Cardelli et al.(1993a)]{c93a} Cardelli, J~A., Mathis, J.~S., Ebbets, D.~C., \& Savage, B.~D.\ 1993a, \apjl, 402, L17
\bibitem[Cardelli \& Meyer(1997)]{cm97} Cardelli, J.~A., \& Meyer, D.~M.\ 1997, \apjl, 477, L57
\bibitem[Cardelli et al.(1996)]{c96} Cardelli, J.~A., Meyer, D.~M., Jura, M., \& Savage, B.~D.\ 1996, \apj, 467, 334
\bibitem[Cardelli et al.(1991a)]{c91a} Cardelli, J.~A., Savage, B.~D., Bruhweiler, F.~C., et al.\ 1991a, \apjl, 377, L57
\bibitem[Cardelli et al.(1991b)]{c91b} Cardelli, J.~A., Savage, B.~D., \& Ebbets, D.~C.\ 1991b, \apjl, 383, L23
\bibitem[Cardelli et al.(1994)]{css94} Cardelli, J.~A., Sofia, U.~J., Savage, B.~D., Keenan, F.~P., \& Dufton, P.~L.\ 1994, \apjl, 420, L29
\bibitem[Cartledge et al.(2004)]{c04} Cartledge, S.~I.~B., Lauroesch, J.~T., Meyer, D.~M., \& Sofia, U.~J.\ 2004, \apj, 613, 1037
\bibitem[Cartledge et al.(2006)]{c06} Cartledge, S.~I.~B., Lauroesch, J.~T., Meyer, D.~M., \& Sofia, U.~J.\ 2006, \apj, 641, 327
\bibitem[Cartledge et al.(2008)]{c08} Cartledge, S.~I.~B., Lauroesch, J.~T., Meyer, D.~M., Sofia, U.~J., \& Clayton, G.~C.\ 2008, \apj, 687, 1043
\bibitem[Cartledge et al.(2003)]{c03} Cartledge, S.~I.~B., Meyer, D.~M., \& Lauroesch, J.~T.\ 2003, \apj, 597, 408
\bibitem[Cartledge et al.(2001)]{c01} Cartledge, S.~I.~B., Meyer, D.~M., Lauroesch, J.~T., \& Sofia, U.~J.\ 2001, \apj, 562, 394
\bibitem[Cashman et al.(2017)]{c17} Cashman, F.~H., Kulkarni, V.~P., Kisielius, R., Ferland, G.~J., \& Bogdanovich, P.\ 2017, \apjs, 230, 8
\bibitem[Chiappini et al.(1997)]{c97} Chiappini, C., Matteucci, F., \& Gratton, R.\ 1997, \apj, 477, 765
\bibitem[de Avillez \& Mac Low(2002)]{dm02} de Avillez, M.~A., \& Mac Low, M.-M.\ 2002, \apj, 581, 1047
\bibitem[De Cia et al.(2021)]{dc21} De Cia, A., Jenkins, E.~B., Fox, A.~J., et al.\ 2021, Nature, 597, 206
\bibitem[De Cia et al.(2022)]{dc22} De Cia, A., Jenkins, E.~B., Fox, A.~J., et al.\ 2022, Nature, 605, E8
\bibitem[Diplas \& Savage(1994)]{ds94} Diplas, A., \& Savage, B.~D.\ 1994, \apjs, 93, 211
\bibitem[Donor et al.(2020)]{d20} Donor, J., Frinchaboy, P.~M., Cunha, K., et al.\ 2020, \aj, 159, 199
\bibitem[Dwek \& Scalo(1980)]{ds80} Dwek, E., \& Scalo, J.~M.\ 1980, \apj, 239, 193
\bibitem[Edmunds(1975)]{e75} Edmunds, M.~G.\ 1975, \apss, 32, 483
\bibitem[Esteban et al.(2022)]{e22} Esteban, C., M{\'e}ndez-Delgado, J.~E., Garc{\'i}a-Rojas, J., \& Arellano-C{\'o}rdova, K.~Z.\ 2022, \apj, 931, 92
\bibitem[Federman et al.(2007)]{f07} Federman, S.~R., Brown, M., Torok, S., et al.\ 2007, \apj, 660, 919
\bibitem[Federman et al.(2003)]{f03} Federman, S.~R., Lambert, D.~L., Sheffer, Y., et al.\ 2003, \apj, 591, 986
\bibitem[Fran{\c{c}}ois et al.(2004)]{f04} Fran{\c{c}}ois, P., Matteucci, F. Cayrel, R., et al.\ 2004, \aap, 421, 613
\bibitem[Heidarian et al.(2017)]{h17} Heidarian, N., Irving, R.~E., Federman, S.~R., et al.\ 2017, JPhB, 50, 155007
\bibitem[Heidarian et al.(2015)]{h15} Heidarian, N., Irving, R.~E., Ritchey, A.~M., et al.\ 2015, \apj, 808, 112
\bibitem[Hobbs et al.(1993)]{h93} Hobbs, L.~M., Welty, D.~E., Morton, D.~C., Spitzer, L., \& York, D.~G.\ 1993, \apj, 411, 750
\bibitem[Jenkins(1996)]{j96} Jenkins, E.~B.\ 1996, \apj, 471, 292
\bibitem[Jenkins(2009)]{j09} Jenkins, E.~B.\ 2009, \apj, 700, 1299
\bibitem[Jenkins(2013)]{j13} Jenkins, E.\ 2013, in Proc.~of The Life Cycle of Dust in the Universe: Observations, Theory, and Laboratory Experiments, ed.~A.~Andersen et al.~(Trieste: SISSA), 5
\bibitem[Jenkins(2019)]{j19} Jenkins, E.~B.\ 2019, \apj, 872, 55
\bibitem[Jenkins et al.(1986)]{j86} Jenkins, E.~B., Savage, B.~D., \& Spitzer, L.\ 1986, \apj, 301, 355
\bibitem[Jenkins \& Tripp(2006)]{jt06} Jenkins, E.~B., \& Tripp, T.~M.\ 2006, \apj, 637, 548
\bibitem[Jensen et al.(2007)]{j07} Jensen, A.~G., Rachford, B.~L., \& Snow, T.~P.\ 2007, \apj, 654, 955
\bibitem[Jensen \& Snow(2007)]{js07} Jensen, A.~G., \& Snow, T.~P.\ 2007, \apj, 669, 378
\bibitem[Jones et al.(1994)]{j94} Jones, A.~P., Tielens, A.~G.~G.~M., Hollenbach, D.~J., \& McKee, C.~F.\ 1994, ApJ, 433, 797
\bibitem[Jura \& York(1978)]{jy78} Jura, M., \& York, D.~G.\ 1978, \apj, 219, 861
\bibitem[Kisielius et al.(2015)]{k15} Kisielius, R., Kulkarni, V.~P., Ferland, G.~J., et al.\ 2015, \apj, 804, 76
\bibitem[Knauth et al.(2003)]{k03} Knauth, D.~C., Andersson, B-G, McCandliss, S.~R., \& Moos, H.~W.\ 2003, \apjl, 596, L51
\bibitem[Knauth et al.(2006)]{k06} Knauth, D.~C., Meyer, D.~M., \& Lauroesch, J.~T.\ 2006, \apjl, 647, L115
\bibitem[Lambert et al.(1998)]{l98} Lambert, D.~L., Sheffer, Y., Federman, S.~R., et al.\ 1998, \apj, 494, 614
\bibitem[Lodders(2003)]{l03} Lodders, K.\ 2003, \apj, 591, 1220
\bibitem[Luck(2018)]{l18} Luck, R.~E.\ 2018, \aj, 156, 171
\bibitem[Lundberg et al.(2016)]{l16} Lundberg, H., Hartman, H., Engstrom, L., et al.\ 2016, \mnras, 460, 356
\bibitem[M{\'e}ndez-Delgado et al.(2022)]{md22} M{\'e}ndez-Delgado, J.~E., Amayo, A., Arellano-C{\'o}rdova, K.~Z., et al.\ 2022, \mnras, 510, 4436
\bibitem[Meyer et al.(1997)]{m97} Meyer, D.~M., Cardelli, J.~A., \& Sofia, U.~J.\ 1997, \apjl, 490, L103
\bibitem[Meyer et al.(1998)]{m98} Meyer, D.~M., Jura, M., \& Cardelli, J.~A.\ 1998, \apj, 493, 222
\bibitem[Miller et al.(2007)]{m07} Miller, A., Lauroesch, J.~T., Sofia, U.~J., Cartledge, S.~I.~B., \& Meyer, D,~M.\ 2007, \apj, 659, 441
\bibitem[Morton(2000)]{m00} Morton, D.~C.\ 2000, \apjs, 130, 403
\bibitem[Morton(2003)]{m03} Morton, D.~C.\ 2003, \apjs, 149, 205
\bibitem[Nieva \& Przybilla(2012)]{np12} Nieva, M.-F., \& Przybilla, N.\ 2012, \aap, 539, A143
\bibitem[Oliver \& Hibbert(2013)]{oh13} Oliver, P., \& Hibbert, A.\ 2013, ADNDT, 99, 459
\bibitem[Petit et al.(2015)]{p15} Petit, A.~C., Krumholz, M.~R., Goldbaum, N.~J., \& Forbes, J.~C.\ 2015, \mnras, 449, 2588
\bibitem[Press et al.(2007)]{p07} Press, W. H., Teukolsky, S. A., Vetterling, W. T., \& Flannery, B. P.\ 2007, Numerical Recipes, The Art of Scientific Computing (3rd ed.; Cambridge: Cambridge Univ. Press)
\bibitem[Rachford et al.(2009)]{r09} Rachford, B.~L., Snow, T.~P., Destree, J.~D., et al.\ 2009, \apjs, 180, 125
\bibitem[Rachford et al.(2002)]{r02} Rachford, B.~L., Snow, T.~P., Tumlinson, J., et al.\ 2002, \apj, 577, 221
\bibitem[Ritchey et al.(2018)]{r18} Ritchey, A.~M., Federman, S.~R., \& Lambert, D.~L.\ 2018, \apjs, 236, 36
\bibitem[Ritchey et al.(2011)]{r11} Ritchey, A.~M., Federman, S.~R., Sheffer, Y., \& Lambert, D.~L.\ 2011, \apj, 728, 70
\bibitem[Ritchey et al.(2023)]{r23} Ritchey, A.~M., Brown, J.~M., Federman, S.~R., \& Sonnentrucker, P.\ 2023, \apj, in press [arXiv:2301.09727]
\bibitem[Roth \& Blades(1995)]{rb95} Roth, K.~C., \& Blades, J.~C.\ 1995, \apjl, 445, L95
\bibitem[Roy \& Kunth(1995)]{rk95} Roy, J.-R., \& Kunth, D.\ 1995, \aap, 294, 432
\bibitem[Savage et al.(1992)]{s92} Savage, B.~D., Cardelli, J.~A., \& Sofia, U.~J.\ 1992, \apj, 401, 706
\bibitem[Savage \& Sembach(1991)]{ss91} Savage, B.~D., \& Sembach, K.~R.\ 1991, \apj, 379, 245
\bibitem[Savage \& Sembach(1996)]{ss96b} Savage, B.~D., \& Sembach, K.~R.\ 1996, \araa, 34, 279
\bibitem[Schectman et al.(2005)]{s05} Schectman, R.~M., Federman, S.~R., Brown, M., et al.\ 2005, \apj, 621, 1159
\bibitem[Seab \& Shull(1983)]{ss83} Seab, C.~G., \& Shull, J.~M.\ 1983, \apj, 275, 652
\bibitem[Sembach \& Savage(1996)]{ss96a} Sembach, K.~R., \& Savage, B.~D.\ 1996, \apj, 457, 211
\bibitem[Sheffer et al.(2008)]{s08} Sheffer, Y., Rogers, M., Federman, S.~R., et al.\ 2008, \apj, 687, 1075
\bibitem[Shull et al.(1977)]{s77} Shull, J.~M., York, D.~G., \& Hobbs, L.~M.\ 1977, \apjl, 211, L139
\bibitem[Slavin et al.(2015)]{s15} Slavin, J.~D., Dwek, E., \& Jones, A.~P.\ 2015, ApJ, 803, 7
\bibitem[Snow et al.(2002)]{srf02} Snow, T.~P., Rachford, B.~L., \& Figoski, L.\ 2002, \apj, 573, 662
\bibitem[Snow et al.(2000)]{s00} Snow, T.~P., Rachford, B.~L., Tumlinson, J., et al.\ 2000, \apjl, 538, L65
\bibitem[Sofia et al.(1997)]{s97} Sofia, U.~J., Cardelli, J.~A., Guerin, K.~P., \& Meyer, D.~M.\ 1997, \apjl, 482, L105
\bibitem[Sofia et al.(1994)]{s94} Sofia, U.~J., Cardelli, J.~A., \& Savage, B.~D.\ 1994, \apj, 430, 650
\bibitem[Sofia et al.(2004)]{s04} Sofia, U.~J., Lauroesch, J.~T., Meyer, D.~M., \& Cartledge, S.~I.~B.\ 2004, \apj, 605, 272
\bibitem[Sofia et al.(1999)]{s99} Sofia, U.~J., Meyer, D.~M., \& Cardelli, J.~A.\ 1999, \apjl, 522, L137
\bibitem[Sonnentrucker et al.(2002)]{s02} Sonnentrucker, P., Friedman, S.~D., Welty, D.~E., York, D.~G., \& Snow, T.~P.\ 2002, \apj, 576, 241
\bibitem[Sonnentrucker et al.(2003)]{s03} Sonnentrucker, P., Friedman, S.~D., Welty, D.~E., York, D.~G., \& Snow, T.~P.\ 2003, \apj, 596, 350
\bibitem[Sonnentrucker et al.(2007)]{s07} Sonnentrucker, P., Welty, D.~E., Thorburn, J.~A., \& York, D.~G.\ 2007, \apjs, 168, 58
\bibitem[Spitzer(1985)]{s85} Spitzer, L.\ 1985, \apjl, 290, L21
\bibitem[Timmes et al.(1995)]{t95} Timmes, F.~X., Woosley, S.~E., \& Weaver, T.~A.\ 1995, \apjs, 98, 617
\bibitem[Toner \& Hibbert(2005)]{th05} Toner, A., \& Hibbert, A.\ 2005, \mnras, 361, 673
\bibitem[Welty(2007)]{w07} Welty, D.~E.\ 2007, \apj, 668, 1012
\bibitem[Welty \& Crowther(2010)]{wc10} Welty, D.~E., \& Crowther, P.~A.\ 2010, \mnras, 404, 1321
\bibitem[Welty et al.(1999)]{w99} Welty, D.~E., Hobbs, L.~M., Lauroesch, J.~T., et al.\ 1999, \apjs, 124, 465
\bibitem[Welty et al.(1995)]{w95} Welty, D.~E., Hobbs, L.~M., Lauroesch, J.~T., Morton, D.~C., \& York, D.~G.\ 1995, \apjl, 449, L135
\bibitem[Welty et al.(2020)]{w20} Welty, D.~E., Sonnentrucker, P., Snow, T.~P., \& York, D.~G.\ 2020, \apj, 897, 36
\bibitem[Yang \& Krumholz(2012)]{yk12} Yang, C.-C., \& Krumholz, M.\ 2012, \apj, 758, 48
\end{thebibliography}
\end{document}